\documentclass[fleqn,usenatbib]{mnras}

\usepackage{newtxtext,newtxmath}

\usepackage[T1]{fontenc}

\DeclareRobustCommand{\VAN}[3]{#2}
\let\VANthebibliography\thebibliography
\def\thebibliography{\DeclareRobustCommand{\VAN}[3]{##3}\VANthebibliography}

\usepackage{graphicx}	
\usepackage{amsmath}	

\title[HD~145718]{Scattering and sublimation: a multi-scale view of $\mu$m-sized dust in the inclined disc of HD~145718}

\author[C.~L. Davies et al.]{
Claire L. Davies,$^{1}$\thanks{E-mail: c.davies3@exeter.ac.uk}
Evan A. Rich,$^{2}$ 
Tim J. Harries,$^{1}$ 
John D. Monnier,$^{2}$ 
Anna S. E. Laws,$^{1}$ 
\newauthor
Sean M. Andrews,$^{3}$ 
Jaehan Bae,$^{4,5}$ 
David J. Wilner,$^{3}$ 
Narsireddy Anugu,$^{6,7}$ 
Jacob Ennis,$^{2}$ 
\newauthor
Tyler Gardner,$^{2}$ 
Stefan Kraus,$^{1}$ 
Aaron Labdon,$^{1,8}$ 
Jean-Baptiste le Bouquin,$^{2,9}$ 
Cyprien Lanthermann,$^{6}$ 
\newauthor
Gail H. Schaefer,$^{6}$ 
Benjamin R. Setterholm,$^{2}$ 
Theo ten Brummelaar,$^{6}$ 
and the G-LIGHTS collaboration
\\
$^{1}$Astrophysics Group, Department of Physics and Astronomy, University of Exeter, Stocker Road, Exeter, EX4 4QL, UK\\
$^{2}$Astronomy Department, University of Michigan, Ann Arbor, MI 48109, USA\\
$^{3}$Center for Astrophysics \textbar\ Harvard \& Smithsonian, 60 Garden Street, Cambridge, MA 02138, USA\\
$^{4}$Earth and Planets Laboratory, Carnegie Institution for Science, 5241 Broad Branch Road NW, Washington, DC 20015, USA\\
$^{5}$Department of Astronomy, University of Florida, Gainesville, FL 32611, USA\\
$^{6}$Steward Observatory, Department of Astronomy, University of Arizona, Tucson, USA\\
$^{7}$The CHARA Array of Georgia State University, Mount Wilson Observatory, Mount Wilson, CA 91023, USA\\
$^{8}$European Southern Observatory, Casilla 19001, Santiago 19, Chile\\
$^{9}$Institut de Planetologie et d'Astrophysique de Grenoble, Grenoble 38058, France
}

\date{Accepted XXX. Received YYY; in original form ZZZ}

\pubyear{2021}

\begin{document}
\label{firstpage}
\pagerange{\pageref{firstpage}--\pageref{lastpage}}
\maketitle

\begin{abstract}
We present multi-instrument observations of the disc around the Herbig~Ae star, HD~145718, employing geometric and Monte Carlo radiative transfer models to explore the disc orientation, the vertical and radial extent of the near infrared (NIR) scattering surface, and the properties of the dust in the disc surface and sublimation rim.
The disc appears inclined at $67-71^{\circ}$, with position angle, PA\,$=-1.0-0.6^{\circ}$, consistent with previous estimates.
The NIR scattering surface extends out to $\sim75\,$au and we infer an aspect ratio, $h_{\rm{scat}}(r)/r\sim0.24$ in $J$-band; $\sim0.22$ in $H$-band.
Our GPI images and VLTI+CHARA NIR interferometry suggest that the disc surface layers are populated by grains $\gtrsim \lambda/2\pi$ in size, indicating these grains are aerodynamically supported against settling and/or the density of smaller grains is relatively low.
We demonstrate that our geometric analysis provides a reasonable assessment of the height of the NIR scattering surface at the outer edge of the disc and, if the inclination can be independently constrained, has the potential to probe the flaring exponent of the scattering surface in similarly inclined ($i\gtrsim70^{\circ}$) discs.
In re-evaluating HD~145718's stellar properties, we found that the object's dimming events - previously characterised as UX~Or and dipper variability - are consistent with dust occultation by grains larger, on average, than found in the ISM.
This occulting dust likely originates close to the inferred dust sublimation radius at $0.17\,$au.
\end{abstract}

\begin{keywords}
accretion discs -- radiative transfer -- techniques: high angular resolution -- circumstellar matter -- stars: individual: HD~145718 -- stars: formation
\end{keywords}



\section{Introduction}
The planet formation process requires sub-$\mu$m sized particles, typical of the interstellar medium (ISM), to grow by >12 orders of magnitude to produce planetesimals and planets. Moreover, given the relatively short lifetimes of protoplanetary discs ($\sim3-11\,$Myr, \citealt{Ribas15}), such growth has to be highly efficient. The process is complicated - with grain evolution involving coagulation, vertical settling, radial drift, and fragmentation - and dependent on the local structure of the disc and the properties of the dust therein \citep{Testi14}.

Highly inclined ($i\gtrsim70^{\circ}$), dust rich protoplanetary discs provide unique opportunities to study vertical and radial disc structure. In particular, the optically thick nature of protoplanetary discs across the optical/near-infrared (NIR) allows observations at these wavelengths to directly trace the disc surface layers. The dust in these regions is expected to be dominated by sub-$\mu$m sized grains and be well-coupled to the gas. Meanwhile, large grains are expected to preferentially settle towards the disc midplane \citep{Dubrulle95}. 

Observational evidence for such vertical stratification is seen in the wavelength dependence of the vertical extent of near-edge-on discs \citep[e.g.][]{Glauser08, Duchene10, Villenave19, Wolff21}. However in apparent contrast, the presence of ``large'' aggregates (radius, $a\gtrsim \lambda/2\pi$, where $\lambda$ is the observed wavelength) in the surface layers of protoplanetary discs has been inferred from (i) asymmetric brightness patterns in polarised differential imaging data \citep[e.g.][]{Mulders13, Stolker16a, Stolker16b, Avenhaus18, Garufi20} and (ii) the colours and polarisation of stars exhibiting photometric variability attributable to occultation by circumstellar dust (namely the ``dippers'' \citealt{Bouvier99, Stauffer15, Bredall20} and UX~Ors \citealt{Huang19}). Numerical simulations and lab experiments show that such large, porous dust grains may be key to overcoming the bouncing barrier \citep{Wada11, Kothe13, Brisset17} and the radial drift barrier \citep{Okuzumi12, Kataoka13} to dust grain growth.

Bright, relatively close-by ($d\lesssim300\,$pc) young stellar objects (YSOs) permit the detailed study of disc structure using 8\,m-class telescopes and infrared (IR) and millimetre (mm) interferometers. Here, we focus on HD~145718 (common aliases include PDS~80 and V718~Sco), an intermediate mass YSO ($\sim1.5-3\,\rm{M}_{\odot}$) - i.e. a Herbig Ae star - in the Upper Scorpius (USco) association \citep{Rizzuto11, Pecaut12, Luhman18}. In particular, we combine new and archival observations of the dusty circumstellar environment around HD~145718 probing sub-au to tens of au scales. We combine these observations to constrain the nature of the dust grains in the surface layers of the innermost and outermost disc regions and assess whether circumstellar dust obscuration is likely responsible for the photometric variability observed in this object.

Our paper is organised as follows. Section~\ref{sec:hd145718} provides an overview of previous studies involving HD~145718 and our knowledge of its circumstellar environment to-date. Section~\ref{sec:observations} describes our Gemini Planet Imager (GPI) observations of HD~145718, conducted as part of the Gemini Large Imaging with GPI Herbig/T-Tauri Survey (G-LIGHTS; \citealt[][Rich et al. 2021b, \textit{in prep}]{Monnier17, monnier2019, laws2020}). Sections~\ref{subsec:GRAV_PION} and \ref{sec:extantData} describe our complementary CHARA/MIRC-X NIR interferometric observations, VLTI/PIONIER and VLTI/GRAVITY archival NIR interferometric datasets, and archival multi-band, multi-epoch photometry and IR spectroscopy. Our combination of NIR interferometric and polarised scattered light imaging probes sub-$\mu$m- to $\mu$m-sized dust grains on multiple angular scales. We first employ a simple geometric model to explore the disc orientation and extent. The methodology and results of this part of our investigation are presented in Section~\ref{sec:analytical}. We further build on this in Section~\ref{sec:RTmodeling} using full Monte Carlo radiative transfer models to simultaneously model the GPI images, NIR interferometry, and spectral energy distribution. This includes a re-evaluation of HD~145718's stellar luminosity, radius and visual extinction in Section~\ref{sec:starParam}. We discuss our results in the context of HD~145718's photometric variability in Section~\ref{sec:geometry}, assess the robustness of our geometric modelling in Section~\ref{subsec:surface}, and summarise our findings in Section~\ref{sec:summary}.

\section{HD 145718}\label{sec:hd145718}
Previous studies of HD~145718 have reported the existence of an inclined ($i\gtrsim70^{\circ}$; \citealt{Guimaraes06, Garufi18, Gravity19, Ansdell20}), gas-rich \citep{Dent05, Ansdell20} and dust-rich \citep[e.g.][]{GregorioHetem92, Oudmaijer92, Friedemann96}, \citet{Meeus01} Group II disc \citep{Keller08}. \citet{Dent05} obtained a marginal $J=3-2$ $^{12}$CO detection towards HD~145718 and (accounting for the different stellar distances adopted between their study and ours - see Section~\ref{sec:starParam}) estimated the gaseous disc extends out to $70\pm35\,$au.

The star itself has a spectral type of A5 \citep{Carmona10}. Its identification as photometrically variable saw it classified as an eclipsing binary throughout the 1900s. However, by comparing their radial velocities to earlier measurements by \citet{Carmona10}, \citet{Ripepi15} found no evidence of multiplicity in the system. Adaptive optics imaging has also ruled out the presence of companions at $20$-$200\,$milliarcsecond (mas) separations down to $\Delta L'=2.6$--$4.8\,$mag \citep{Ansdell20}. More recently, HD~145718's photometric variability has been re-attributed to inherent stellar variability ($\delta$ Scuti-type pulsations; \citealt{Ripepi15}) and obscuration by circumstellar dust (dipper and UX~Ori variability; \citealt{Poxon15, Ansdell18, Cody18, Rebull18}). Temporal variations like those seen in the blue-shifted portion of HD~145718's H$\alpha$ line profile (compare, for example, the line profiles presented in \citealt{Vieira03}, \citealt{Carmona10}, \citealt{Ripepi15}, and \citealt{Wichittanakom20}) and higher order Balmer series lines \citep{Guimaraes06} can also be attributed to accretion and outflow processes in YSOs \citep[e.g.][]{Muzerolle04} and are likely also linked to the dipper/UX~Ori variability.

The evolutionary status of HD~145718 has been debated in the literature and isochronal age estimates for the object vary from $5.7$ to $20\,$Myr \citep{Alecian13, Fairlamb15, Vioque18, Arun19, Wichittanakom20}. However, estimating the age of individual disc-hosting, photometrically variable young stars using isochrone fitting is fraught with difficulty \citep[e.g.][]{Davies14}. Strong evidence for HD~145718 being pre-main-sequence rather than an evolved star is found in the $p$-mode frequencies of the object's $\delta$ Scuti-type pulsations: the highest $p$-mode frequency observed - which scales linearly with stellar age \citep{Zwintz14} - is consistent with those of disc-less, non-accreting USco members with isochronal ages of $\sim10\,$Myr \citep{Ripepi15}.

\section{Observations and supplementary archival data}
\subsection{GPI data}\label{sec:observations}
$J$- and $H$-band polarimetry mode observations of HD~145718 with GPI \citep{Macintosh14}, situated at the Gemini South telescope, were obtained on 2018Jun07 and 2018Jun08, respectively (program ID GS-2018A-LP-12). 
The 32 frames, each with 2 co-adds, were observed with exposure times of $29\,$s (total integration time $=1862\,$s per waveband).
Between each frame, the half-wave plate was rotated between $0{^\circ}$, $22.5{^\circ}$, $45{^\circ}$, and $67.5{^\circ}$, creating eight independent sequences with four equally spaced half-wave plate angles. 
Additionally, $J$- and $H$- band coronagraphs were used with focal plane diameters of of $0.184''$ and $0.246''$, respectively.

To reduce the data, we used the GPI Data Reduction Pipeline (DRP) version 1.5 \citep{perrin2014, maire2010}, with modifications to the flux calibration of the polarised images and the removal of stellar and instrumental polarisation (see \citealt{monnier2019}, \citealt{laws2020}, and Rich et al. 2021b, \emph{in prep}, for details). 
In summary, the GPI DRP was used to subtract dark background, extract the polarisation spots, correct for bad pixels, remove microphonics noise, flat-field the frames using a low-frequency flat, and measure the star position using a radon transformation of the satellite spots\footnote{These so-called satellite spots are created by diffractive elements in the pupil plane of the GPI instrument. Their radial structure point to the star's location behind the coronographic mask.}. 
This resulted in 32 polarised images: four polarised images within each of the eight polarisation sets. 
Each of the four polarised images were combined together using the double-differencing technique, creating eight Stokes cubes containing $I$, $Q$, and $U$ images. 
The polarisation sets were rotated so that the top of the image pointed north and the stellar and instrumental polarisation were removed (Appendix~\ref{app:starpolsub}).
We then projected the eight polarisation sets of $Q$ and $U$ (oriented with North up; East left) to local $Q_\phi$ and $U_\phi$, based on the stellar position determined above. Specifically, for a pixel grid of ($X$, $Y$) coordinates with centre pixel ($X_{0}$,$Y_{0}$) and coordinate position angle, $\gamma$:
\begin{equation}
    Q_{\phi}=-U\sin \left(2\phi \right) - Q\cos \left(2\phi \right)
\end{equation}
\begin{equation}
    U_{\phi}=Q\sin \left(2\phi \right) - U\cos \left(2\phi \right)
\end{equation}
where $\phi$ is the polar angle:
\begin{equation}
    \phi = \tan^{-1} \left( \frac{Y-Y_{0}}{X - X_{0}} \right) + \gamma
\end{equation}
(see Appendix~A of \citealt{monnier2019}). The 8 polarisation sets were then averaged together to create the combined $I$, $Q_\phi$, and $U_\phi$ images.

The satellite spots in the polarised images were used in the flux calibration. The 32 polarised images were averaged together to increase the signal-to-noise of the satellite spots, as described in \citet{laws2020}. We used only the second order and only the first order satellite spots for the $J$- and $H$-band observations, respectively. Using 2MASS photometry \citep[][Table~\ref{tab:phot}]{Cutri2003ya}, we measured $J$- and $H$-band flux scaling factors of $3.387\pm0.75\,$mJy\,arcsec$^{-1}$\,/(ADU/sec/coadd) and $2.492\pm0.37\,$mJy\,arcsec$^{-1}$\,/(ADU/sec/coadd), respectively. 

\begin{figure*}
  \centering
	\includegraphics[trim=1.5cm 0.9cm 1.5cm 0.2cm, clip=true,width=0.44\textwidth]{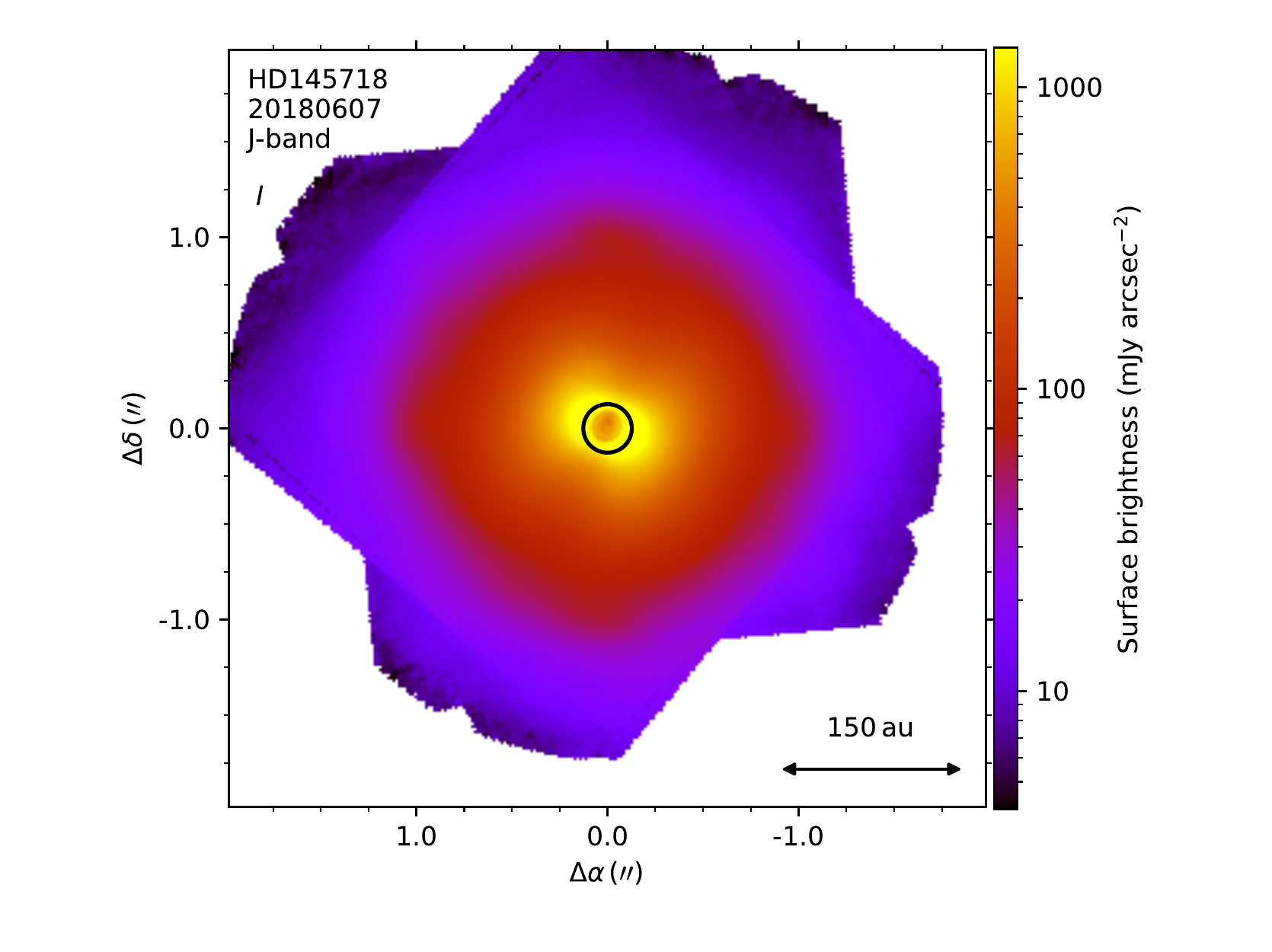}
	\includegraphics[trim=1.5cm 0.9cm 1.5cm 0.2cm, clip=true,width=0.44\textwidth]{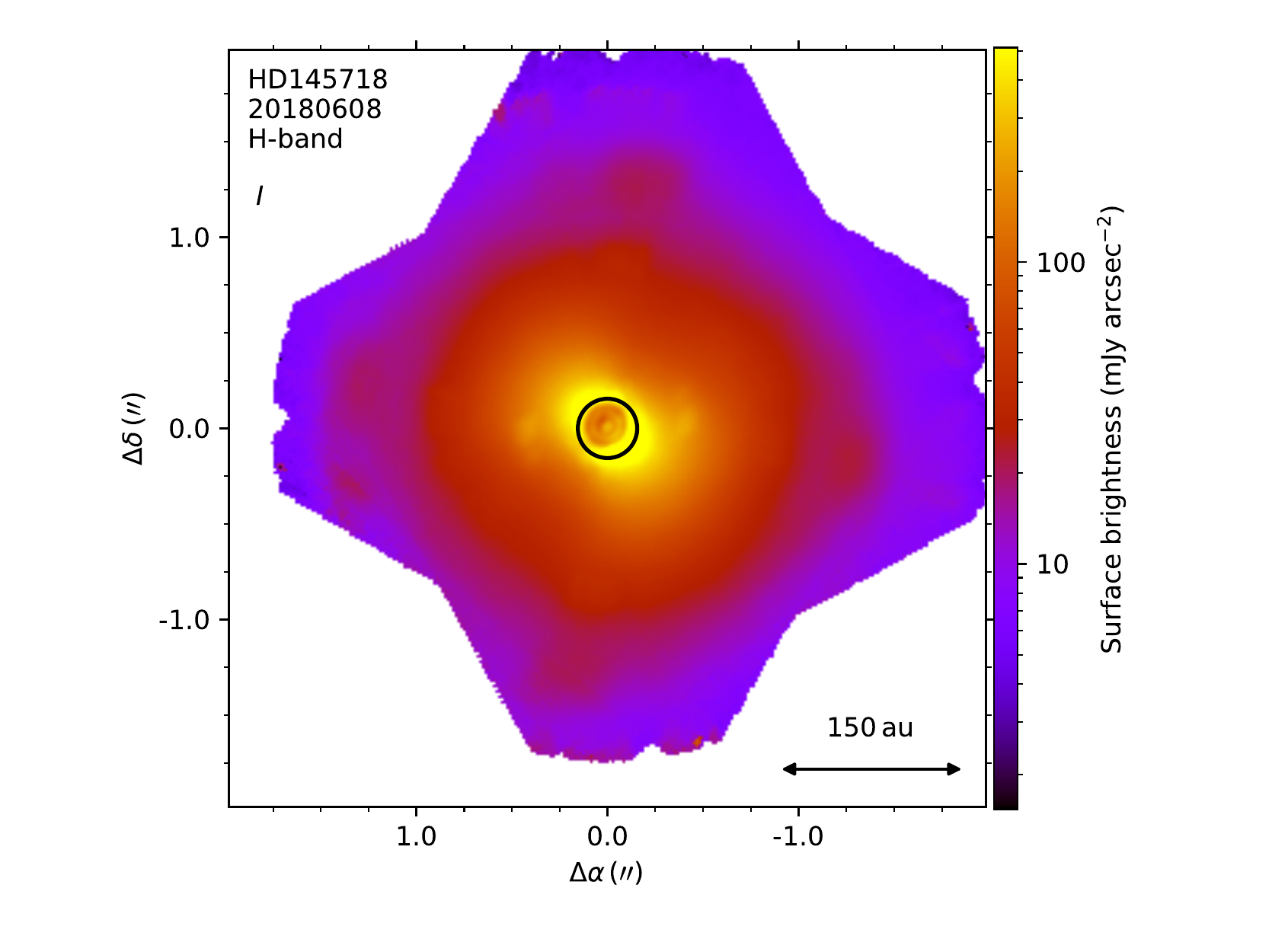}\\
	\includegraphics[trim=1.5cm 0.9cm 1.5cm 0.2cm, clip=true,width=0.44\textwidth]{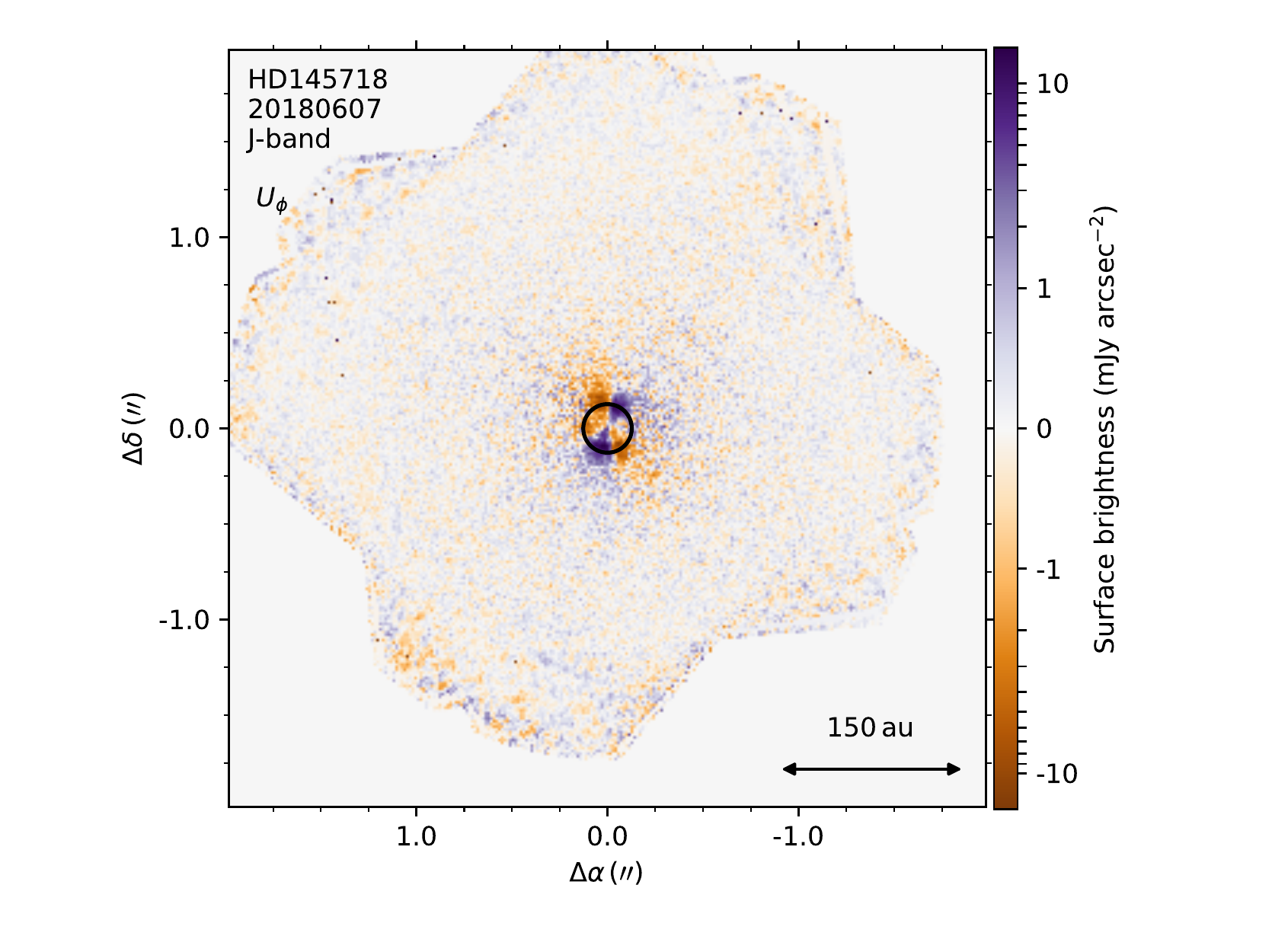}
	\includegraphics[trim=1.5cm 0.9cm 1.5cm 0.2cm, clip=true,width=0.44\textwidth]{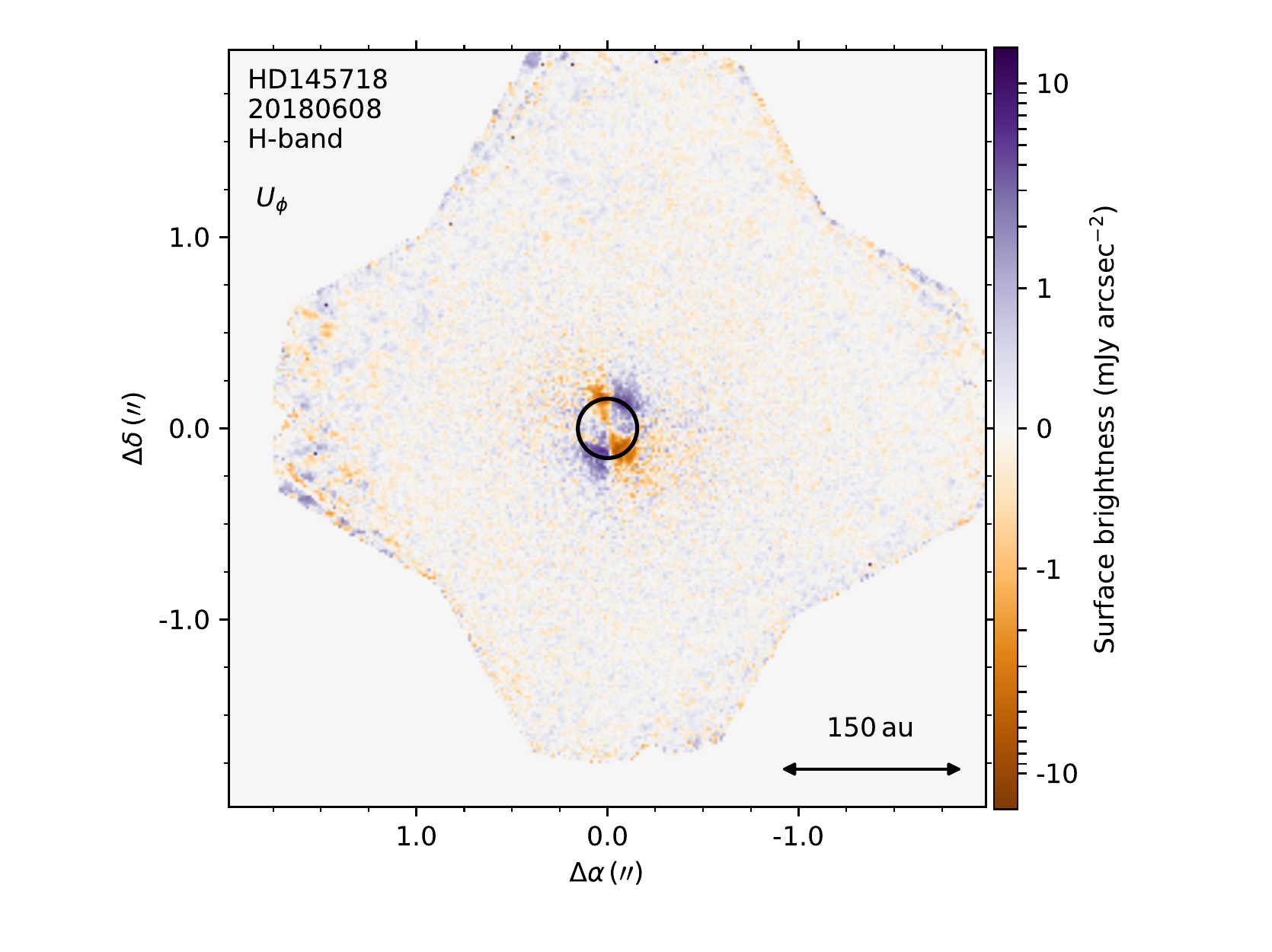}\\
	\includegraphics[trim=1.5cm 0.9cm 1.4cm 0.2cm, clip=true,width=0.44\textwidth]{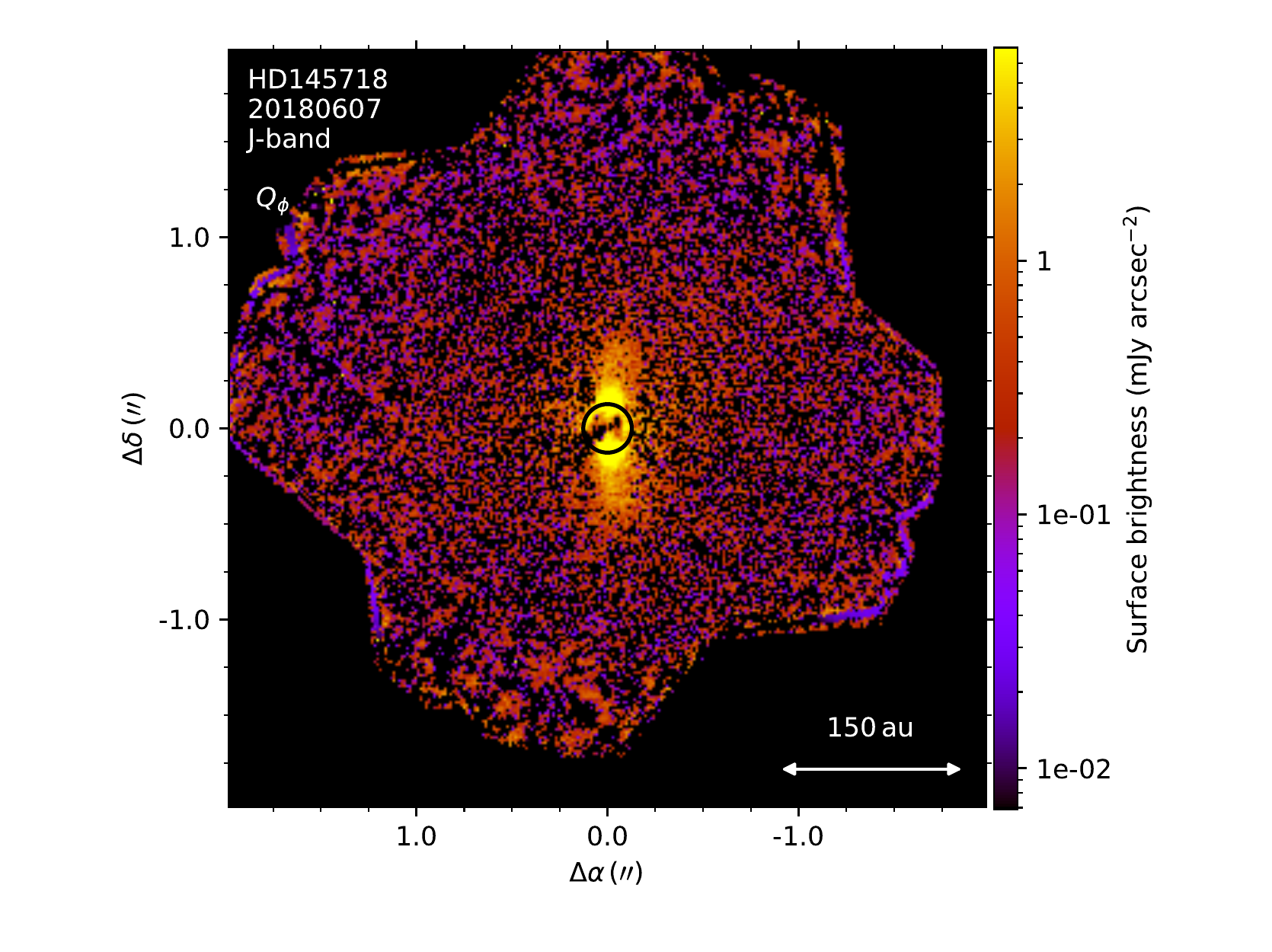}
	\includegraphics[trim=1.6cm 0.9cm 1.3cm 0.2cm, clip=true,width=0.44\textwidth]{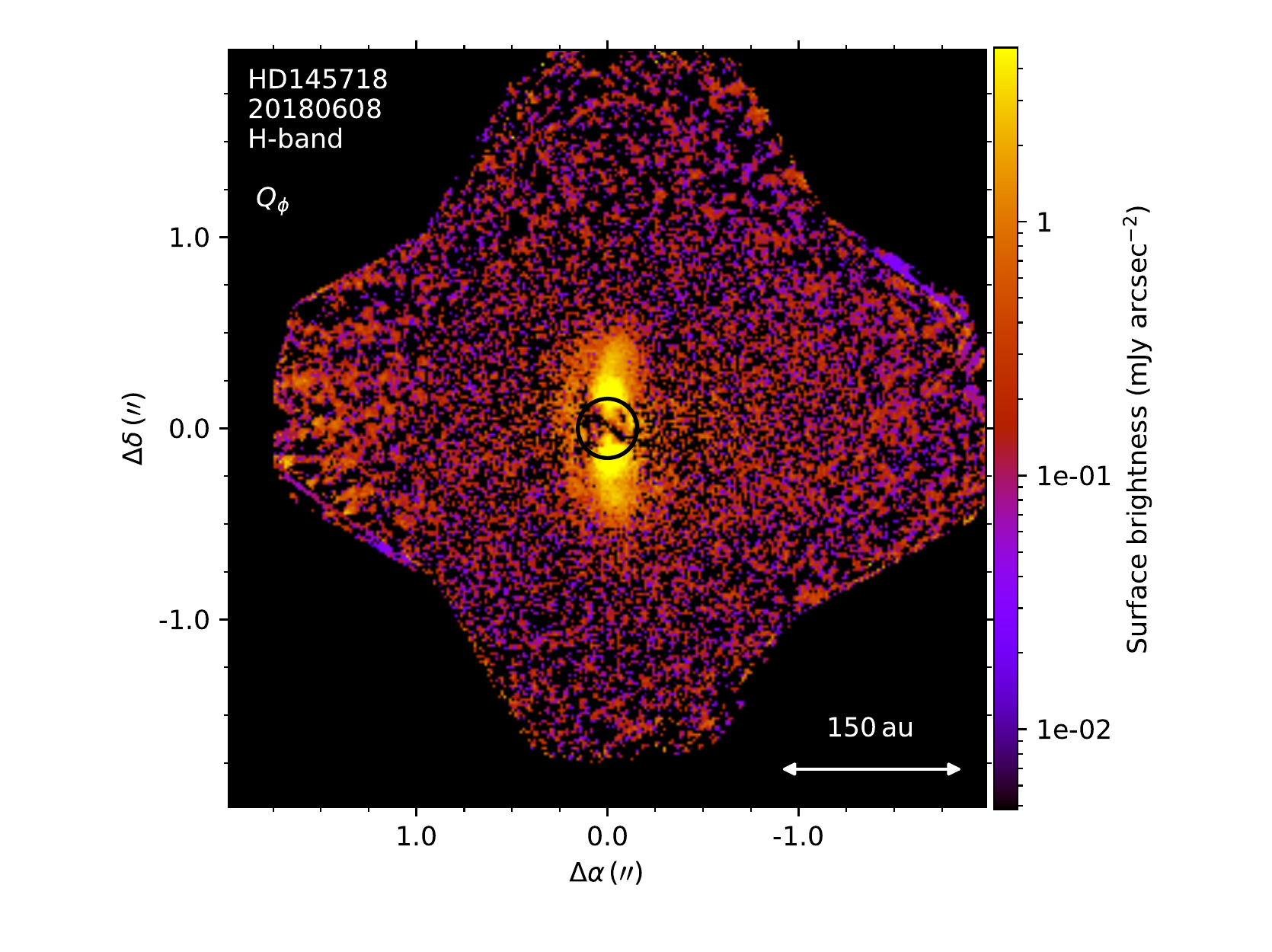}
    \caption{GPI images: total intensity, $I$ (top row), radial Stokes $U_{\phi}$ (middle row) and $Q_{\phi}$ (bottom row). The left (right) column shows the $J$-band ($H$-band) images (north is up; east is left). The physical scale is shown in the bottom right corner of all the images. The IWA of the coronographic mask (radii of $\sim9\,$ and $\sim11\,$pixels for the $J$- and $H$-band, respectively) is indicated by a black ring in the centre of each image.}
    \label{fig:gpi_obs}
\end{figure*}

\begin{figure}
    \centering
	\includegraphics[trim=2.5cm 0.9cm 1.8cm 0.2cm, clip=true,width=0.44\textwidth]{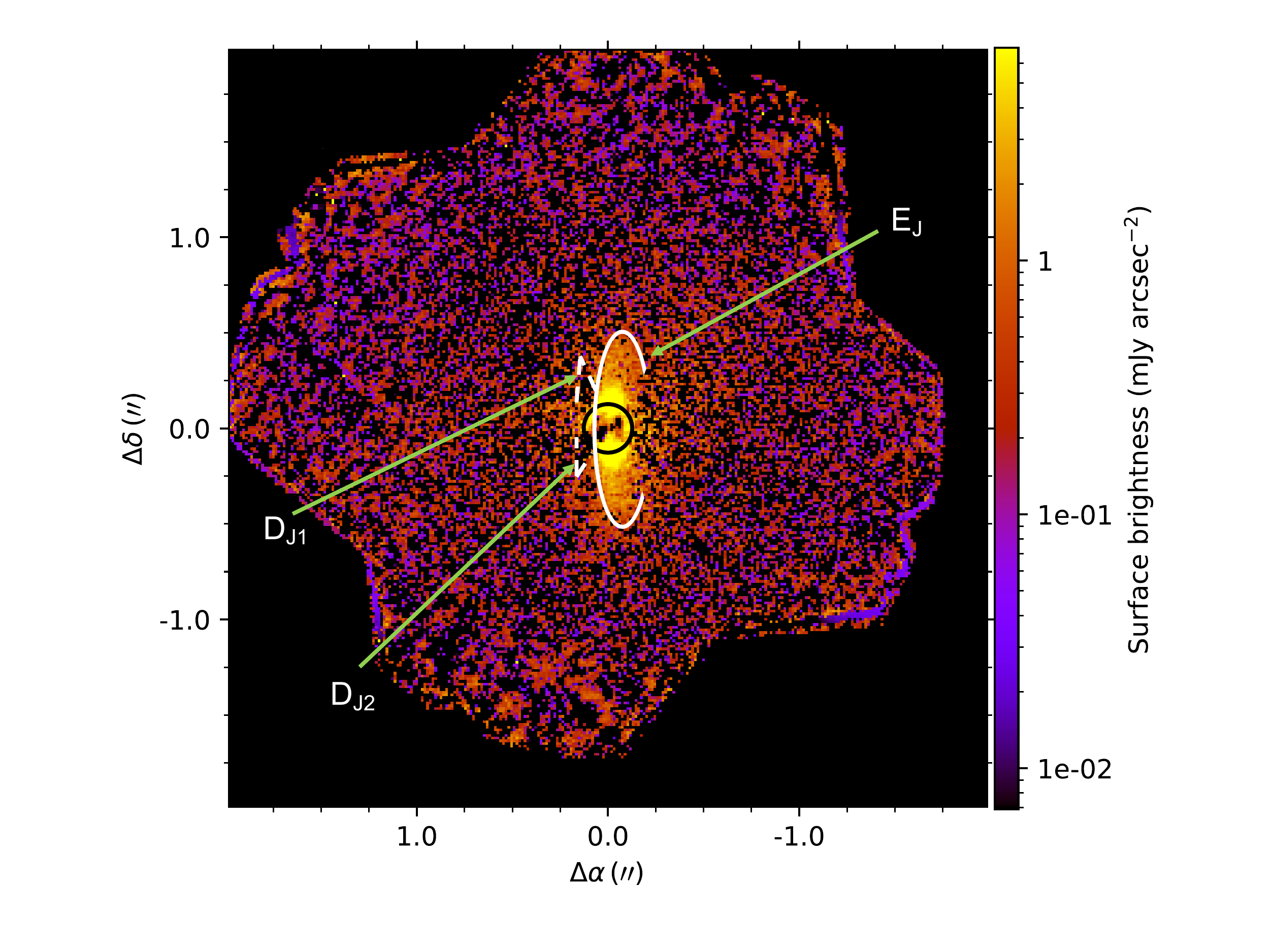}
	\includegraphics[trim=2.5cm 0.9cm 1.8cm 0.2cm, clip=true,width=0.44\textwidth]{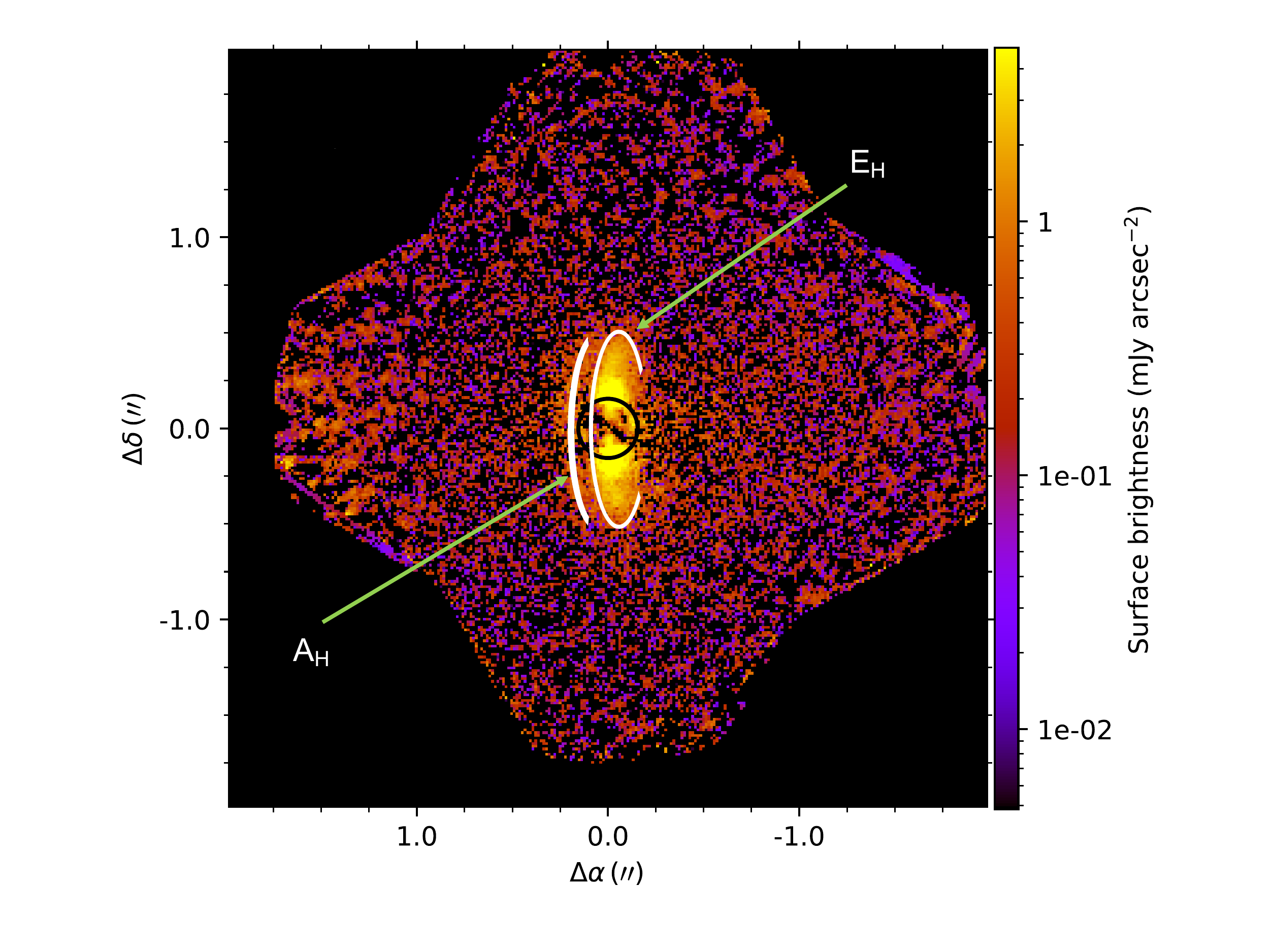}
    \caption{Annotated $J$- (top) and $H$-band (bottom) Stokes $Q_{\phi}$ images highlighting the main features $E_{\rm{J}}$, $D_{\rm{J1}}$, $D_{\rm{J2}}$, $E_{\rm{H}}$, and $A_{\rm{H}}$ (see Section~\ref{subsec:inspect} for details). The IWA of the coronographic mask is indicated by the black ring in the centre of the image.}
    \label{fig:gpi_annotated}
\end{figure}

\subsubsection{Visual inspection of flux-calibrated images}\label{subsec:inspect}
Our flux-calibrated Stokes $I$, $U_\phi$, and $Q_\phi$ images of HD~145718 are shown in the top, middle and bottom rows of Figure~\ref{fig:gpi_obs}, respectively. Figure~\ref{fig:gpi_annotated} highlights the main features in the Q$_\phi$ images. The black circle at the centre of each image indicates the size of the inner working angle (IWA) of the coronographic mask (radii of $\sim9\,$ and $\sim11\,$pixels for the $J$- and $H$-band images, respectively; Rich et al. 2021b, \textit{in prep}). 

Both the $J$- and $H$-band $Q_{\phi}$ images feature a central ellipse (marked $E_{\rm{J}}$ and $E_{\rm{H}}$, respectively, in Figure~\ref{fig:gpi_annotated}), elongated along a north--south direction. The $U_{\phi}$ images feature positive flux to the north-west and south-east and negative flux in the north-east and south-west. The brightest regions (negative and positive flux) predominantly extend along a north--south direction with minimal extension to the east and west of the IWA. Taken together, these $Q_{\phi}$ and $U_{\phi}$ features are consistent with the presence of an inclined circumstellar disc around HD~145718, with a major axis position angle, PA$\,\approx0^{\circ}$. This is consistent with previous results from K-band and mm continuum interferometry: PA$\,=2\pm2^{\circ}$ \citep{Gravity19} and PA$\,=1\pm1^{\circ}$ \citep{Ansdell20}, respectively.

The major axes of $E_{\rm{J}}$ and $E_{\rm{H}}$ appear offset to the west of the image centre. While we cannot directly measure how well the coronograph is centered on the star in HD~145718, we can use other targets in our G-LIGHTS sample with companion detections within the field-of-view of GPI to comment on the significance of this apparent offset. For example, in the eight cycles we observed for G-LIGHTS target HD~50138, we estimate a standard deviation for the centroid of its companion to be $\sim0.22\,$pixels ($\sim3.1\,$mas), much smaller than the offset we observe. Thus, we do not expect the coronograph centering to be a large contributing factor to the offset we observe for HD~145718. Instead, the offset of the ellipse centre from the image centre is likely symptomatic of a flared disc structure where the ellipse traces the open face of the disc towards the observer. 

The $H$-band $Q_{\phi}$ image features a bright arc to the east of $E_{\rm{H}}$ (marked $A_{\rm{H}}$ in Figure~\ref{fig:gpi_obs}). This likely traces scattering events close to the outer edge of the surface of the disc facing away from the observer. The drop in surface brightness between $E_{\rm{H}}$ and $A_{\rm{H}}$ is likely a result of the opaque disc midplane. The $J$-band $Q_{\phi}$ image lacks a similar arc feature. Instead, two dark features ($Q_{\phi}$ flux significantly below the background level) are observed immediately to the east of the coronographic mask (marked $D_{\rm{J1}}$ and $D_{\rm{J2}}$ in Figure~\ref{fig:gpi_obs}). These may also be attributable to the opaque disc midplane or may be artifacts of imperfect stellar polarisation subtraction (Appendix~\ref{app:starpolsub}).

\subsection{Complementary near-infrared interferometry}\label{subsec:GRAV_PION}
\begin{figure}
    \centering
    \includegraphics[trim=0.8cm 0.0cm 1.1cm 0.0cm, clip=true,width=0.23\textwidth]{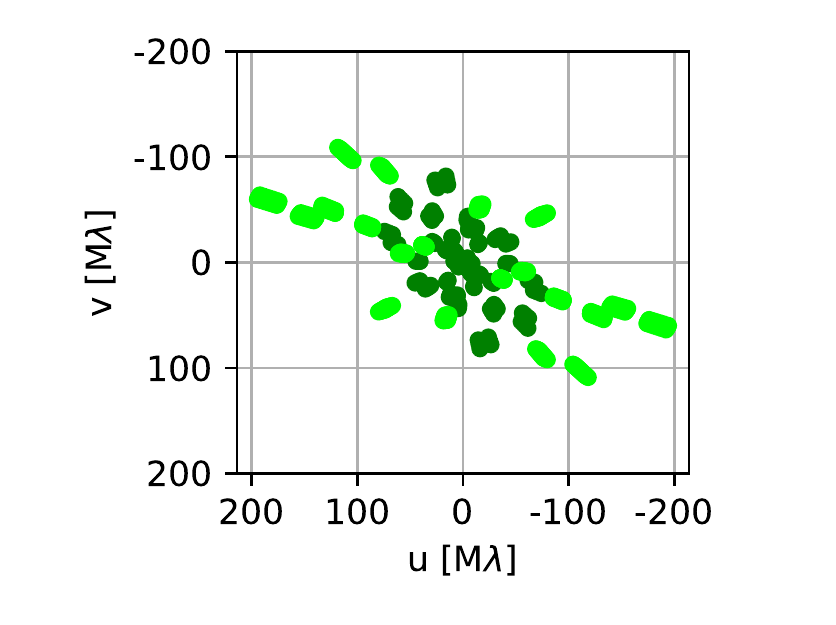}
    \includegraphics[trim=0.8cm 0.0cm 1.1cm 0.0cm, clip=true,width=0.23\textwidth]{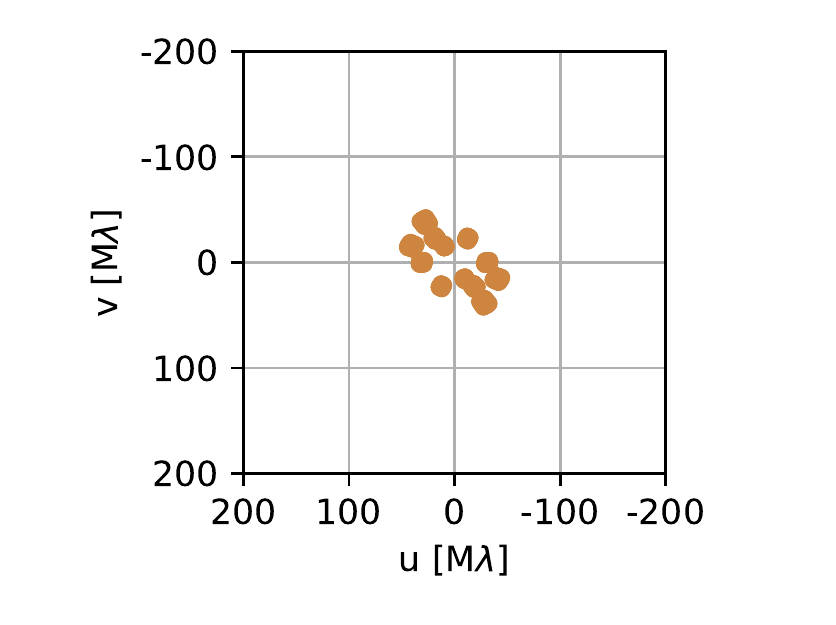}\\
    \caption{($u,v$)-plane coverage of the interferometry. Left: $H$-band VLTI/PIONIER (dark green data points) and CHARA/MIRC-X (lime green data points). Right: $K$-band VLTI/GRAVITY (brown data points). North is up; east is left.}
    \label{fig:uv}
\end{figure}

Fully reduced and calibrated VLTI/PIONIER data were retrieved from the Optical Interferometry Database (OIDB). These probe $H$-band emission from HD~145718 on mas scales. Details of the reduction and calibration procedure are provided in \citet{Lazareff2017pv}. 
$K$-band NIR interferometric data, obtained using VLTI/GRAVITY and originally published in \citet{Gravity19}, were retrieved from the European Southern Observatory archive. The data were reduced and calibrated using the GRAVITY pipeline (version 1.1.2) using default settings. We restrict our analysis to the low spectral dispersion ($R=\Delta \lambda / \lambda \sim30$) GRAVITY fringe tracker (FT) data. The standard star HD~145809 (uniform disc diameter, UDD$\,=0.402\pm0.002\,$mas; \citealt{Bourges17}) was used to estimate the transfer function and calibrate the visibilities and closure phases. The bluest spectral channel of the GRAVITY FT was not used as this is known to be corrupted by the metrology laser operating at $1.908\,\mu$m \citep{Lippa16}.

To probe smaller-scale circumstellar emission, we also obtained a single snapshot observation of HD~145718 using the MIRC-X instrument \citep{Kraus18, Anugu18, Anugu20} of the Centre for High Angular Resolution Astronomy (CHARA) Array on UT date 2021May12. The ($u,v$)-plane coverage of our observations are compared to those of the GRAVITY and PIONIER interferometric datasets in Figure~\ref{fig:uv}. The CHARA Array comprises six $1\,$m-class telescopes arranged in a Y-shaped array. Its operational baselines between $34-330\,$m \citep{tenBrummelaar05} provide $\sim0.5\,$mas resolution\footnote{$\approx\lambda/2B$, with $B$ denoting the baseline length and $\lambda$ the operational wavelength.} across the $H$-band. MIRC-X is capable of combining light from all six CHARA telescopes. However, fiber injection issues on one beam associated with a telescope focus problem caused by the mounting mechanism of the primary mirror limited our observations to the five-telescope configuration E1-W2-W1-S2-E2. We used the PRISM\,50 spectral mode which provides five spectral channels across the $H$-band ($\lambda\sim1.4-1.7\,\mu$m). The data were reduced using pipeline version 1.3.5\footnote{https://gitlab.chara.gsu.edu/lebouquj/mircx\_pipeline.git.}, described in \citet{Anugu20}. We applied the bispectrum bias correction, set the number of coherent coadds to 10, and adopted a flux threshold of 10. Otherwise, we adopted default reduction settings. Standard stars HD~145965 (UDD\,$=0.209\pm0.005\,$mas; \citealt{Bourges17}) and HD~139487 (UDD\,$=0.305\pm0.009\,$mas; \citealt{Bourges17}) were observed either side of HD~145718 in a CAL-SCI-CAL concatenation. These data were inspected for signatures of binarity but none were found. They were used to estimate the transfer function to calibrate the visibilities and closure phases. 

\subsubsection{Inspection of the NIR interferometry}
The CHARA/MIRC-X interferometry was inspected for (i) consistency with the VLTI/PIONIER data and (ii) signatures of binarity. The shorter baseline MIRC-X visibilities showed good consistency with those obtained by PIONIER. However, the different angular scales and position angles probed by the two datasets - together with the variations in brightness that HD~145718 is known to exhibit - make direct comparison of the data difficult. The MIRC-X closure phases are consistent with zero, indicating that the underlying brightness distribution is centro-symmetric. The VLTI data were known to show a similar lack of deviation from centro-symmetry \citep{Lazareff2017pv, Gravity19}. We thus find no indication of binarity and restrict our analysis in Section~\ref{sec:RTmodeling} to the visibilities.  

\subsection{Archival multi-wavelength photometry and infrared spectroscopy}\label{sec:extantData}
Complementary multi-wavelength archival photometry and flux measurements were retrieved using the Spectral Energy Distribution Builder for Young Stars (SEDBYS, \citealt{Davies21}). The WISE W4-band magnitude was flagged and removed due to its discrepant low flux and we added the $1.3\,$mm ALMA flux reported in \citet{Garufi18} to the collated dataset. The full list of flux-calibrated photometry, together with their references, are provided in Appendix~\ref{appen:phot}. Flux-calibrated \textit{Spitzer} Infrared Spectrograph (IRS; \citealt{Houck04}) Short-Low and Long-High module spectra were also retrieved from the Infrared Science Archive (IRSA). Details regarding the reduction of these data are provided in \citet{Keller08}. 

\section{Geometric modeling of the GPI images}\label{sec:analytical}
We fit elliptical ring models to isophotes of surface brightness, $S_{\nu}$, tracing the $E_{\rm{J}}$ and $E_{\rm{H}}$ features in the $Q_{\phi}$ images. These allow us to assess the radial and vertical extents, as well as the orientation, of the disc around HD~145718 prior to the more computationally expensive and time-consuming radiative transfer modelling (Section~\ref{sec:RTmodeling}). Our elliptical ring model was prescribed as a circular ring of radius, $r$, inclined by angle, $i$, rotated through position angle, PA\footnote{Position angles are those of the disc major axis, measured East of North.}, and translated in right ascension, RA, and declination, Dec, by coordinates ($\delta\,$RA, $\delta\,$Dec) from the image centre. Assuming our observations trace light scattered by dust close to the disc surface, we expect the centres of $E_{\rm{J}}$ and $E_{\rm{H}}$ to be offset from the image centre along a vector which, when projected onto the sky, lies perpendicular to the disc PA. Thus, we can relate PA, $\delta\,$RA, and $\delta\,$Dec to the height, $h_{\rm{scat}}(r)$, of the scattering surface above the disc midplane at radius $r$:
\begin{equation}\label{eq:h_shift}
    \rm{PA} = \tan^{-1}\left(\frac{\delta\,\rm{RA}}{\delta\,\rm{Dec}} \right) + \frac{\pi}{2},
\end{equation}
and 
\begin{equation}\label{eq:h_from_coords}
    h_{\rm{scat}}(r) = d\left( \left(\delta\,\rm{RA}\right)^{2} + \left(\delta\,\rm{Dec}\right)^{2} \right)^{1/2},
\end{equation}
where $d$ is the stellar distance. We follow \citet{Vioque18} and adopt $d=152.5^{+3.2}_{-3.0}\,$pc, corresponding to the inverse of the Gaia data release (DR) 2 parallax\footnote{This is within the range of the \citet{Bailer18} estimate of $151.9\pm1.9\,$pc (which accounts for the nonlinear nature of the parallax--distance transformation), based on the Gaia DR2 parallax, and the Gaia early DR3 inverse parallax estimate of $154.7\pm0.5\,$pc \citep{Gaia21, Lindegren21}.} (\citealt{Gaia16, Gaia18}). Our elliptical ring model is then fully prescribed using four parameters: $r$, $i$, PA, and $h_{\rm{scat}}(r)$. 

Before extracting the $S_{\nu}$ isophotes, we first masked the $Q_{\phi}$ images to exclude the central pixels within the IWA of the coronograph. The pixel coordinates of the $S_{\nu}$ isophotes tracing the $E_{\rm{J}}$ and $E_{\rm{H}}$ features were then isolated from the full list returned by the {\tt contour} function of {\tt matplotlib.pyplot} \citep{Hunter07}. The western side of each ellipse-tracing $S_{\nu}$ isophote deviated from an elliptical shape, likely due to the combined effects of (i) the relatively narrow vertical extent of the scattering surface compared to the east--west extent of the coronographic mask; and (ii) the scattering phase function of the dust grains in the disc resulting in a lower back-scattered than forward-scattered flux \citep[e.g.][see Section~\ref{subsec:torusGrid}]{Stolker16b, Tazaki19}. Meanwhile, the eastern side of each $S_{\nu}$ isophote was shaped by features $D_{\rm{J1}}$, $D_{\rm{J2}}$ and $A_{\rm{H}}$. To isolate the ellipse-tracing portion of each $S_{\nu}$ isophote, the collated coordinate arrays were inspected by-eye and cuts were applied to the horizontal and vertical pixel coordinates. Additionally, isophotes with $S_{\nu}>1.25\,\rm{mJy}\,\rm{arcsec}^{-2}$ ($J$-band) or $S_{\nu}>1.85\,\rm{mJy}\,\rm{arcsec}^{-2}$ ($H$-band) and $S_{\nu}<0.70\,\rm{mJy}\,\rm{arcsec}^{-2}$ ($J$-band) or $S_{\nu}<0.80\,\rm{mJy}\,\rm{arcsec}^{-2}$ ($H$-band) were not used as they did not sufficiently trace the apexes of the ellipse. Limiting the range of $S_{\nu}$ isophotes used in this way results in a limited range of disc radii (and therefore $h_{\rm{scat}}(r)$) being explored. 

\subsection{Inferred disc geometry and potential flaring of the disc scattering surface}
\begin{table}
    \centering
    \caption{Results of our geometric modelling of the $S_{\nu}$ isophotes of the $J$- and $H$-band $Q_{\phi}$ images. Column 1: isophote surface brightness; columns 2, 3 and 4: elliptical ring radius, inclination, and position angle; column 5: height of the scattering surface above the disc midplane at radius $r$.} \label{tab:qphiScaleHeight}
     \begin{tabular}{ccccc}
    \hline
    $S_{\nu}$ & $r$ & $i$ & PA & $h_{\rm{scat}}(r)$ \\ 
    (mJy\,/arcsec$^{2}$)  & (au) & ($^{\circ}$) & ($^{\circ}$) & (au)\\ 
    (1) & (2) & (3) & (4) & (5) \\ 
    \hline
    \multicolumn{5}{c}{$J$-band} \\
    \hline
    $0.70$ & $71.4^{+0.6}_{-0.6}$ & $68.1^{+0.6}_{-0.7}$ &  $-0.7^{+0.3}_{-0.4}$ &  $9.5^{+0.4}_{-0.4}$  \\ 
    $0.75$ & $71.0^{+0.6}_{-0.6}$ & $68.3^{+0.7}_{-0.7}$ &  $-0.5^{+0.3}_{-0.4}$ &  $9.6^{+0.5}_{-0.4}$ \\ 
    $0.80$ & $70.7^{+0.6}_{-0.6}$ & $68.2^{+0.7}_{-0.7}$ &  $-0.6^{+0.3}_{-0.4}$ &  $9.8^{+0.5}_{-0.5}$ \\ 
    $0.85$ & $69.5^{+0.6}_{-0.6}$ & $67.1^{+0.7}_{-0.7}$ &  $-0.1^{+0.3}_{-0.4}$ &  $10.3^{+0.5}_{-0.5}$ \\ 
    $0.90$ & $69.1^{+0.7}_{-0.6}$ & $66.9^{+0.8}_{-0.9}$ &  $-0.6^{+0.4}_{-0.5}$ &  $10.8^{+0.6}_{-0.5}$ \\ 
    $0.95$ & $66.9^{+0.6}_{-0.6}$ & $67.7^{+0.8}_{-0.9}$ &  $-0.1^{+0.3}_{-0.4}$ &  $9.2^{+0.6}_{-0.5}$ \\ 
    $1.00$ & $66.6^{+0.6}_{-0.6}$ & $68.4^{+0.8}_{-0.9}$ &  $0.3^{+0.3}_{-0.3}$ &  $9.0^{+0.6}_{-0.6}$ \\ 
    $1.05$ & $66.0^{+0.6}_{-0.6}$ & $69.0^{+0.8}_{-0.8}$ &  $0.6^{+0.3}_{-0.3}$ &  $8.5^{+0.5}_{-0.5}$ \\ 
    $1.10$ & $65.4^{+0.6}_{-0.6}$ & $68.6^{+0.8}_{-0.9}$ &  $0.2^{+0.3}_{-0.4}$ &  $8.8^{+0.5}_{-0.5}$  \\ 
    $1.15$ & $64.7^{+0.6}_{-0.6}$ & $68.6^{+0.8}_{-0.9}$ &  $0.2^{+0.3}_{-0.3}$ &  $9.1^{+0.5}_{-0.5}$  \\ 
    $1.20$ & $63.6^{+0.5}_{-0.5}$ & $67.8^{+0.8}_{-0.8}$ &  $0.1^{+0.3}_{-0.3}$ &  $9.4^{+0.5}_{-0.5}$  \\ 
    $1.25$ & $62.9^{+0.5}_{-0.6}$ & $67.8^{+0.9}_{-1.0}$ &  $0.2^{+0.3}_{-0.4}$ &  $9.4^{+0.7}_{-0.6}$  \\ 
    \hline
    \multicolumn{5}{c}{$H$-band} \\
    \hline
    $0.80$ & $71.6^{+0.5}_{-0.5}$ &  $67.8^{+0.6}_{-0.7}$ &  $-0.1^{+0.3}_{-0.3}$ &  $8.0^{+0.5}_{-0.4}$ \\ 
    $0.85$ &  $71.3^{+0.4}_{-0.4}$ &  $68.6^{+0.5}_{-0.5}$ &  $-0.8^{+0.2}_{-0.3}$ &  $8.3^{+0.3}_{-0.3}$ \\ 
    $0.90$ &  $70.4^{+0.4}_{-0.4}$ &  $69.0^{+0.5}_{-0.5}$ &  $-0.7^{+0.2}_{-0.3}$ &  $7.7^{+0.3}_{-0.3}$ \\ 
    $0.95$ &  $70.0^{+0.5}_{-0.5}$ &  $68.6^{+0.6}_{-0.6}$ &  $-1.0^{+0.3}_{-0.3}$ &  $8.0^{+0.4}_{-0.4}$ \\ 
    $1.00$ &  $69.5^{+0.5}_{-0.5}$ &  $68.5^{+0.6}_{-0.7}$ &  $-1.0^{+0.3}_{-0.4}$ &  $8.3^{+0.4}_{-0.4}$ \\ 
    $1.05$ &  $69.4^{+0.4}_{-0.4}$ &  $69.6^{+0.5}_{-0.5}$ &  $-0.6^{+0.2}_{-0.3}$ &  $7.8^{+0.4}_{-0.4}$ \\ 
    $1.10$ &  $68.3^{+0.4}_{-0.4}$ &  $69.4^{+0.5}_{-0.5}$ &  $-0.3^{+0.2}_{-0.3}$ &  $7.7^{+0.4}_{-0.4}$  \\ 
    $1.15$ &  $68.2^{+0.4}_{-0.4}$ &  $69.4^{+0.5}_{-0.5}$ &  $-0.2^{+0.2}_{-0.3}$ &  $7.7^{+0.4}_{-0.3}$  \\ 
    $1.20$ &  $67.5^{+0.4}_{-0.4}$ &  $69.4^{+0.5}_{-0.5}$ &  $-0.3^{+0.2}_{-0.2}$ &  $7.7^{+0.4}_{-0.3}$  \\ 
    $1.25$ &  $67.5^{+0.4}_{-0.4}$ &  $68.8^{+0.5}_{-0.6}$ &  $-0.5^{+0.2}_{-0.3}$ &  $8.3^{+0.4}_{-0.4}$  \\ 
    $1.30$ &  $66.7^{+0.4}_{-0.4}$ &  $68.3^{+0.6}_{-0.7}$ &  $-0.6^{+0.2}_{-0.3}$ &  $8.5^{+0.4}_{-0.4}$  \\ 
    $1.35$ &  $65.6^{+0.4}_{-0.4}$ &  $69.9^{+0.6}_{-0.6}$ &  $-0.2^{+0.2}_{-0.2}$ &  $6.9^{+0.4}_{-0.4}$ \\ 
    $1.40$ &  $65.4^{+0.4}_{-0.4}$ &  $70.2^{+0.5}_{-0.6}$ &  $-0.1^{+0.2}_{-0.2}$ &  $7.0^{+0.4}_{-0.4}$  \\ 
    $1.45$ &  $65.1^{+0.4}_{-0.4}$ &  $70.7^{+0.5}_{-0.5}$ &  $-0.2^{+0.2}_{-0.2}$ &  $6.8^{+0.4}_{-0.4}$  \\ 
    $1.50$ &  $64.6^{+0.4}_{-0.4}$ &  $70.5^{+0.6}_{-0.6}$ &  $-0.4^{+0.2}_{-0.3}$ &  $7.0^{+0.4}_{-0.4}$  \\ 
    $1.55$ &  $64.5^{+0.4}_{-0.4}$ &  $70.4^{+0.6}_{-0.6}$ &  $-0.5^{+0.2}_{-0.3}$ &  $7.1^{+0.4}_{-0.4}$  \\ 
    $1.60$ &  $63.8^{+0.4}_{-0.4}$ &  $69.6^{+0.7}_{-0.7}$ &  $-0.6^{+0.3}_{-0.3}$ &  $7.7^{+0.5}_{-0.5}$  \\ 
    $1.65$ &  $62.8^{+0.4}_{-0.4}$ &  $69.0^{+0.7}_{-0.7}$ &  $-0.6^{+0.3}_{-0.3}$ &  $7.8^{+0.5}_{-0.5}$  \\ 
    $1.70$ &  $62.3^{+0.4}_{-0.4}$ &  $70.0^{+0.7}_{-0.8}$ &  $-0.7^{+0.3}_{-0.3}$ &  $7.0^{+0.6}_{-0.5}$  \\ 
    $1.75$ &  $62.1^{+0.4}_{-0.4}$ &  $70.0^{+0.7}_{-0.9}$ &  $-0.6^{+0.3}_{-0.3}$ &  $7.0^{+0.7}_{-0.6}$  \\ 
    $1.80$ &  $61.9^{+0.4}_{-0.4}$ &  $70.0^{+0.8}_{-0.9}$ &  $-0.6^{+0.3}_{-0.3}$ &  $7.0^{+0.7}_{-0.6}$  \\ 
    $1.85$ &  $61.6^{+0.5}_{-0.4}$ &  $69.8^{+0.8}_{-0.9}$ &  $-0.7^{+0.3}_{-0.3}$ &  $7.1^{+0.7}_{-0.7}$  \\ 
    \hline
    \end{tabular}
\end{table}

\begin{table*}
    \centering
    \caption{Adopted stellar parameters. The effective temperature (column 2) and surface gravity (column 3) are from \citet{Fairlamb15}. The distance (column 4) is the inverse of the Gaia DR2 parallax \citep{Gaia16, Gaia18}. The visual extinction (column 5), radius (column 6), and luminosity (column 7) were re-evaluated herein using photometry from \citet[][see Appendix~\ref{app:starparam}]{Lazareff2017pv}. } \label{tab:inputParams}
    \begin{tabular}{lccccccc}
    \hline
     & $T_{\rm{eff}}$ & $\log(g)$ & $d$ & $A_{\rm{V}}$ & $R_{\star}$ & $L_{\star}$ \\
     & (K) &  & (pc)  & (mag) & ($R_{\odot})$ & (L$_{\odot}$) \\ 
     & (2) & (3) & (4) & (5) & (6) & (7) \\
    \hline
    HD 145718 & $8000\pm250$ & $4.37\pm0.15$ & $152.5^{+3.2}_{-3.0}$ & $0.89^{+0.34}_{-0.08}$ & $1.97^{+0.12}_{-0.11}$ & $14.3^{+3.9}_{-3.1}$ \\
    \hline
    \end{tabular}
\end{table*}

\begin{figure}
    \centering
    \includegraphics[trim=1.1cm 0.0cm 2.3cm 0.0cm, clip=true,width=0.23\textwidth]{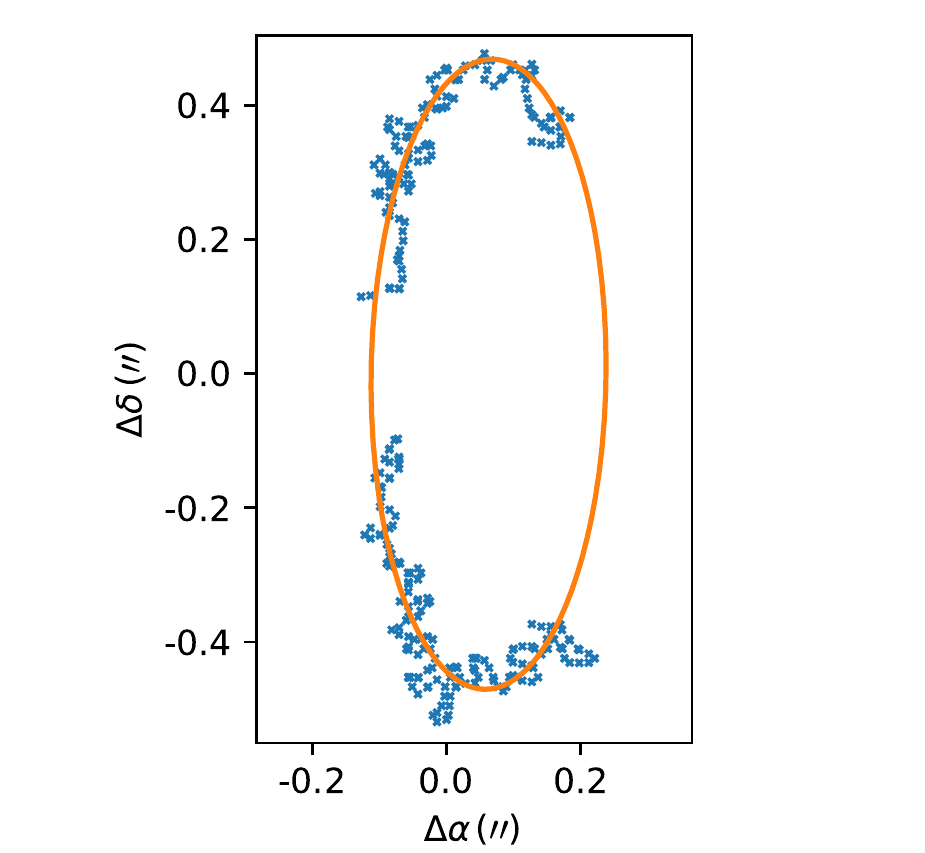}
    \includegraphics[trim=1.0cm 0.0cm 2.3cm 0.0cm, clip=true,width=0.23\textwidth]{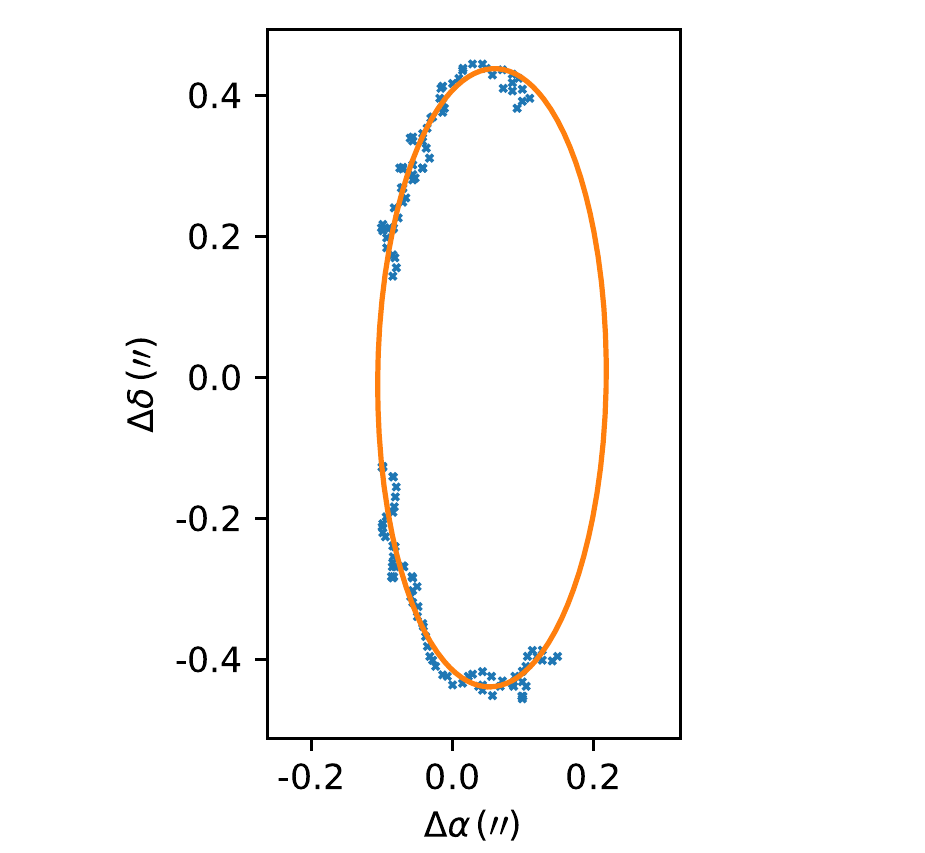}
    \caption{Example isophotes extracted from the $J$- (left) and $H$-band (right) $Q_{\phi}$ images at a surface brightness of $0.7$ and $1.3\,\rm{mJy\,arcsec^{-2}}$, respectively, and their corresponding best fitting elliptical ring model. }
    \label{fig:isophoteFit}
\end{figure}

The {\tt minimize} function of {\tt lmfit} \citep[version 1.0.1;][]{lmfit} was used to fit our elliptical ring model to the data. Specifically, we minimised the radial distance between the cylindrical coordinates of the isophote and the elliptical ring at the same azimuthal angle. Our results are summarised in Table~\ref{tab:qphiScaleHeight} and we show example fits in Fig.~\ref{fig:isophoteFit}. To estimate the uncertainties on our fits, we used the {\tt Minimizer.emcee} package of {\tt lmfit} \citep{emcee, lmfit} to explore the posterior probability distribution of each of the parameters in our model. The values and uncertainties quoted in Table~\ref{tab:qphiScaleHeight} correspond to the median and $1\sigma$ quantiles of these probability distributions. We do not incorporate the uncertainty on the distance (Table~\ref{tab:inputParams}) in our $r$ and $h_{\rm{scat}}(r)$ uncertainties. 

Our fits to the $J$- and $H$-band $S_{\nu}$ isophotes provide broadly consistent results for the disc geometry: $67^{\circ}\lesssim i\lesssim69^{\circ}$ and $-0.7^{\circ}\lesssim\rm{PA_{major}}\lesssim0.6^{\circ}$ in $J$-band; $68^{\circ}\lesssim i\lesssim71^{\circ}$ and $-1.0^{\circ}\lesssim\rm{PA_{major}}\lesssim-0.1^{\circ}$ in $H$-band. As such, the disc appears more inclined and oriented closer to $0^{\circ}$ PA in our GPI images than has previously been inferred from geometric modelling and image reconstruction of PIONIER interferometry \citep{Lazareff2017pv, Kluska20}. Our results are more consistent with findings from Fourier-plane analysis of mm continuum and $K$-band interferometry: $i=70.4\pm1.2^{\circ}$ and PA$\,=1\pm1^{\circ}$ inferred from $1.3\,\mu$m ALMA observations \citep{Ansdell20}; $68^{\circ}<i<72^{\circ}$ and PA$\,=2\pm2^{\circ}$ inferred from GRAVITY observations \citep{Gravity19}. We discuss these results further, particularly in relation to HD~145718's photometric variability, in Section~\ref{sec:geometry}.

The values in Table~\ref{tab:qphiScaleHeight} suggest a scattering surface aspect ratio ($h_{\rm{scat}}(r)/r$) in the range $\sim0.13-0.16$ ($J$-band) and $\sim0.10-0.13$ ($H$-band). This is at the lower end of the range found by e.g. \citet{Avenhaus18} and \citet{Ginski16} who used NIR scattered light images of six protoplanetary discs with concentric ring features to estimate scattering surface aspect ratios, finding $h_{\rm{scat}}(r)/r=0.09-0.25$. The results from this simple isophote fitting procedure also suggest that the $J$-band scattering surface of HD~145718 may be more vertically extended than the $H$-band scattering surface. This is not wholly unexpected as longer wavelengths of light should penetrate deeper into the disc due to the grain size dependent vertical stratification of the disc as a result of vertical settling \citep[e.g.][]{Pinte07, Duchene10}. We examine this in more detail using Monte Carlo radiative transfer models in Section~\ref{subsec:surface}. 

Looking closely at our best-fit $h_{\rm{scat}}(r)$ and $r$ values, the height of the scattering surface also appears to increase with radius suggesting we may be probing the degree of flaring of the disc scattering surface in the outer disc. We explore this further, and assess the robustness of the relatively simplistic fitting procedure we have employed in Section~\ref{subsec:surface}. 

\section{Radiative transfer modeling of the SED, GPI images, and interferometric visibilities}\label{sec:RTmodeling}
We build on our analysis above using Monte Carlo radiative transfer modelling with TORUS \citep{Harries00, Harries19}. This allows us to probe the vertical and radial structure of the disc, and its scattering surface, in a more physically motivated and self-consistent manner. To help constrain our models, we complement our GPI images with new and extant NIR interferometry (Section~\ref{subsec:GRAV_PION}) and multi-wavelength spectro-photometry (Section~\ref{sec:extantData}), allowing us to probe the surface layers of the disc over its full radial extent.

The circumstellar environment of HD~145718 was modelled as a passive disc, illuminated by a central star (see Section~\ref{sec:starParam}), and is built on a two-dimensional, cylindrical adaptive mesh refinement (AMR) grid. The density structure of the gaseous portion of the disc is prescribed following \citet{Shakura73}:
\begin{equation}\label{eq:gas_density}
     \rho_{\rm{gas}}(r, z) = \frac{\Sigma_{\rm{gas}}(r)}{h_{\rm{gas}}(r)\sqrt{2\pi}} \exp\left\{-\frac{1}{2}\left[\frac{z}{h_{\rm{gas}}(r)}\right]^{2}\right\},
\end{equation}
where $r$ and $z$ are the radial distance from the star into the disc and the vertical distance from the disc midplane, respectively. The pressure scale height, $h_{\rm{gas}}(r)$, and surface density, $\Sigma_{\rm{gas}}(r)$, of the gas are prescribed to follow simple radial power laws:
\begin{equation}\label{eq:scaleheight}
   h_{\rm{gas}}(r) = h_{0, \rm{gas}}\left(\frac{r}{r_{0}}\right)^{\beta} 
\end{equation}
and
\begin{equation}\label{eq:surface_density}
    \Sigma_{\rm{gas}}(r) = \Sigma_{0, \rm{gas}}\left(\frac{r}{r_{0}}\right)^{-p}.
\end{equation}
Here, $h_{0, \rm{gas}}$ and $\Sigma_{0, \rm{gas}}$, are the pressure scale height and surface density of the gas, respectively, evaluated at canonical radius, $r_{0}=100\,$au. We keep the surface density power law exponent, $p=1.0$ fixed in all models. 

The \citet{Lucy99} algorithm is used to iteratively solve for radiative equilibrium and dust sublimation. The disc is populated with two populations of dust, comprising ``surface'' and ``settled'' grains (see Section~\ref{sec:dustPrescription}). Dust is added to grid cells whose temperature is cooler than the dust sublimation temperature after the fourth Lucy iteration. Convergence is typically achieved after seven iterations. 

We further used TORUS to generate model SEDs, $4\times4''$ model Stokes $I$, $Q_{\phi}$, and $U_{\phi}$ images at $\lambda=1.25\,\mu$m (J-band) and $1.65\,\mu$m (H-band), and $24\times24\,\rm{mas}$ model total intensity images at $\lambda=1.65\,\mu$m and $2.13\,\mu$m (K-band). We followed the procedure outlined in \citet{Davies18} to extract visibilities from our mas-scale images at the ($u,v$) coordinates of our interferometric dataset (Figure~\ref{fig:uv}). 

\subsection{Stellar parameters}\label{sec:starParam}
The stellar parameters we adopt as input parameters for our TORUS models are listed in Table~\ref{tab:inputParams}. We follow \citet{Vioque18} and adopt the effective temperature, $T_{\rm{eff}}$, and surface gravity, $\log(g)$, estimates from \citet{Fairlamb15}. However, we choose to re-evaluate the stellar luminosity, $L_{\star}$, and visual extinction, $A_{\rm{V}}$, rather than adopt the values in \citet{Vioque18}. Our reasons for this are twofold:
\begin{enumerate}
    \item the \citet{Vieira03} $BVRI$ photometry used by \citet{Vioque18} to estimate $L_{\star}$ trace a fainter epoch than the \citet{Hog2000hg}, \citet{Lazareff2017pv}, and \citet[][]{Gaia18} photometry (see Table~\ref{tab:phot}), suggesting the star may be inherently brighter than the \citet{Vieira03} photometry suggests;
    \item HD~145718's $B-V$ colour is bluer during fainter epochs, consistent with increased scattering during obscuration by circumstellar dust. Indeed, HD~145718 has previously been identified as displaying UX~Ori-type \citep{Poxon15} and dipper variability \citep{Ansdell18, Cody18, Rebull18}. If the dust grains responsible for the occultations are larger, on average, than those in the interstellar medium, the total-to-selective extinction, $R_{\rm{V}}$, may be closer to $5.0$ \citep{Hernandez04} than the value of $3.1$ adopted by \citet{Vioque18}.
\end{enumerate}

Using the brighter epoch $BVRI$ photometry from \citet{Lazareff2017pv}, we follow the methodology outlined in \citet{Fairlamb15} to re-estimate $A_{\rm{V}}$, $R_{\star}$, and $L_{\star}$ (Table~\ref{tab:inputParams}). In doing so, we find that consistent values of $R_{\star}$, and $L_{\star}$ can be used to fit the bright and faint epochs of photometry if $R_{\rm{V}}$ changes from $3.1$ to $5.0$ during dimming events (see Appendix~\ref{app:starparam}). This suggests that dust grains larger, on average, than those found in the ISM are present in the surface layers of the disc. We discuss this further in Section~\ref{subsec:surface}.

\subsection{Disc mass and dust prescription}\label{sec:dustPrescription}
We prescribe populations of ``surface'' and ``settled'' grains in our models, both of which are prescribed as comprising solely of \citet{Draine03} astronomical silicates\footnote{Polyaromatic hydrocarbon (PAH) emission is evident in the IR spectrum of HD~145718 \citep{Keller08} but our assumption is a reasonable approximation as the prominent $10\,\mu$m spectral feature indicates silicate grains are readily abundant in the surface layers of the disc.}. Our settled grains are larger in size and dominate the disc in terms of its mass. Meanwhile, the grains in the disc surface dominate our GPI and interferometric data.

Our surface grain population is assumed to be well-coupled to the gas and therefore follow the vertical and radial density prescriptions in Equations~(\ref{eq:scaleheight}) and (\ref{eq:surface_density}).  They are prescribed to sublimate when they exceed a temperature,
\begin{equation}\label{eqn:Tsub}
    T_{\rm{sub,1}} = G\rho_{\rm{gas}}^{\gamma}\left(r,z\right),
\end{equation}
where $G=2000$ and $\gamma=1.95\times10^{-2}$ \citep{Pollack94}. As the density of disc material is most concentrated around the disc midplane and tapers off at larger scale heights, the dependence of $T_{\rm{sub,1}}$ on $\rho_{\rm{gas}}(r,z)$ results in a curved sublimation rim \citep{Isella05}. How far from the star a grain sublimates also depends on how efficiently it can cool and larger grains cool more efficiently than smaller grains. The location and radial extent of the sublimation rim depends on the size of the largest grains in the mixture \citep{Tannirkulam07} as these will shield smaller grains from incident stellar radiation. We allow our surface grains to range in size between a fixed minimum value, $a_{\rm{min}}=0.01\,\mu$m, and a maximum that is varied between models: $0.14\leq a_{\rm{max}}\leq1.30\,\mu$m. These values of $a_{\rm{max}}$ reflect the range over which an increase in grain size produces an increase in cooling efficiency and associated decrease in sublimation radius \citep[][Davies \& Harries, 2021 \textit{in prep}]{Isella05, Davies20b} and provide a range of scattering phase functions \citep{Stolker16b, Tazaki19}. 

Larger, mm-sized grains are expected to have settled closer to the disc midplane and are therefore absent from the disc surface layers. We restrict the vertical extent of these settled grains to a fraction, $f$, of $h_{\rm{gas}}(r)$, and vary $f$ between models. This increases the density of material in the disc midplane, further affecting the shape and radial extent of the sublimation rim \citep{Tannirkulam07}. How the settling of dust grains larger than a few microns in size influences the location, shape and extent of the sublimation rim has not been well-explored and is beyond the scope of this paper. Instead, after exploring a range of sublimation temperatures for the mm-sized grains, we set the sublimation temperature of the settled grains to a density-independent value of $T_{\rm{sub,2}}=1200\,$K. This ensured that the settled grains were contained within the sublimation rim structure forged by our surface grain population.

Our mm-sized settled grains also dominate the flux at mm-wavelengths. We used the $1.3\,$mm flux, $F_{\rm{\nu}}$, reported in \citet[][see Table~\ref{tab:phot}]{Garufi18}, to estimate a total disc mass (gas+dust),
\begin{equation}\label{eq:discmass}
    M_{\rm{disc}} = \frac{F_{\rm{\nu}} d^{2}}{\kappa_{\nu} B_{\rm{\nu}}\left( T_{\rm{dust}} \right)} = 0.0097\,\rm{M_{\odot}}.
\end{equation}
Here, $B_{\rm{\nu}}\left( T_{\rm{dust}} \right)$ is the blackbody radiation at frequency, $\nu$, for dust at temperature, $T_{\rm{dust}}$. We assumed $T_{\rm{dust}}=20\,$K, and an opacity, $\kappa_{\nu} = 0.1( \nu/10^{12}\,\rm{Hz})^{\beta_{\kappa}} \,\rm{cm^{2}\,g^{-1}}$
with $\beta_{\kappa}=1.0$ \citep{Beckwith90}, which accounts for the adopted 100:1 gas-to-dust ratio. Assuming the density of dust grains follows $n(a)\propto a^{-3.5}$ (where $a$ represents the grain size), the larger, settled grains will contribute $96.7\,$per cent of the dust mass budget (with the smaller, surface grains contributing the remaining $3.3\,$per cent). 

\subsection{TORUS parameter grid exploration}\label{subsec:torusGrid}
\begin{table}
    \centering
    \caption{Best fit disc parameters found from our radiative transfer modelling with TORUS (see Section~\ref{subsec:torusGrid} for the meanings of each of the symbols). Where applicable, physical sizes in au were converted to angular scales using $d=152.5\,$pc (see Section~\ref{sec:analytical}).} \label{tab:RTparams}
    \begin{tabular}{lcc}
    \hline
        Parameter & Values explored & Best model \\
    \hline
        $h_{0, \rm{gas}}$ (au) & $5$, $6$, $7$, $8$, $9$, $10$, $11$ & $10$ \\
        $f$ ($h_{0, \rm{gas}}$) & $0.05$, $0.1$, $0.2$, $0.3$, $0.4$, $0.5$ & $0.1$ \\
        $\beta$ & $ 1.07$, $1.08$, $1.09$, $1.11$, $1.13$, $1.15$, $1.17$ & $1.15$ \\
        $a_{\rm{max}}$ ($\mu$m) & $0.14$, $0.30$, $0.40$, $0.50$, $0.60$, $0.70$, $1.30$ & $0.50$ \\
        $R_{\rm{out}}$ (au) & $70$, $75$, $80$ & $75$ \\
        $i$ ($^{\circ}$) & $48$, $65$, $68$, $70$, $72$, $74$, $76$, $78$, $80$ & $72$ \\
        PA ($^{\circ}$) & $-2$, $0$, $+2$ & $0$ \\
    \hline
    \end{tabular}
\end{table}

We computed a grid of TORUS models in which we varied the gas scale height, $h_{0, \rm{gas}}$; the settling height of the $1\,$mm-sized grains, $f$; the flaring exponent of the gas pressure scale height, $\beta$; the disc outer radius, $R_{\rm{out}}$; the disc orientation ($i$ and PA); and the maximum size of the dust grains in the disc surface layers, $a_{\rm{max}}$ (see Table~\ref{tab:RTparams}). For each model, we visually inspected the fit to the SED, the GPI $Q_{\phi}$ and $U_{\phi}$ images, and the NIR visibilities. 

In assessing the goodness of fit of each model to the SED, the flux across optical to IR wavelengths produced by the model was allowed to range between the bounds of the dereddened and non-dereddened data (black and grey data points in the top panel of Fig~\ref{fig:RTbest_sedOLBI}, respectively). In this way, we assumed that the dereddened observed photometry (using $A_{\rm{V}}=0.89$ and $R_{\rm{V}}=3.1$) provided an upper limit to the allowed model flux and the non-dereddened observed photometry (i.e. $A_{\rm{V}}=0$) provided an extreme lower limit.

In the paragraphs that follow, we briefly discuss the impact of these parameters on the shape of the disc and on the synthesised observables before exploring the model providing the best overall fit in greater detail in the next subsection.

\subsection*{Gas scale height}
Increasing $h_{0, \rm{gas}}$ inflates the vertical extent of the gaseous disc across all radii. This results in increased excess emission in the SED across NIR to mm wavelengths and, in general, brighter $Q_{\phi}$ and $U_{\phi}$ images. The height of the scattering surface is also increased, resulting in a broader gap between the elliptical and arc features in the $Q_{\phi}$ images. The NIR visibilities are sensitive to the contrast between the stellar and circumstellar flux components as well as the shape of the circumstellar NIR emitting region. An increase in $h_{0, \rm{gas}}$ results in a greater fraction of circumstellar flux, decreasing the visibilities at spatial frequencies which at least partially resolve the inner disc rim. Above a certain value of $h_{0, \rm{gas}}$ (which is dependent on the values of $\beta$, $a_{\rm{max}}$, and $i$), the inner disc regions are tall enough to result in direct line-of-sight occultation of the central star (as seen previously for RY~Tau; \citealt{Davies20b}). This decreases the SED flux across optical wavelengths and increases the fraction of total NIR flux emanating from the circumstellar regions in the synthesised image, further lowering the visibilities.

\subsection*{Flaring exponent on the gas pressure scale height}
As $\beta$ is increased between models, the spectral index between $\sim20$ and $100\,\mu$m flattens out while the flux across the near-to-mid infrared decreases. The visibility level on any given baseline increases due to the reduction in circumstellar NIR flux while the $Q_{\phi}$ and $U_{\phi}$ images brighten as more material is distributed to larger radii. 

\subsection*{Settled height of mm grains}
\begin{figure}
    \centering
    \includegraphics[trim=0.7cm 0.1cm 1.2cm 0.0cm, clip=true,width=0.48\textwidth]{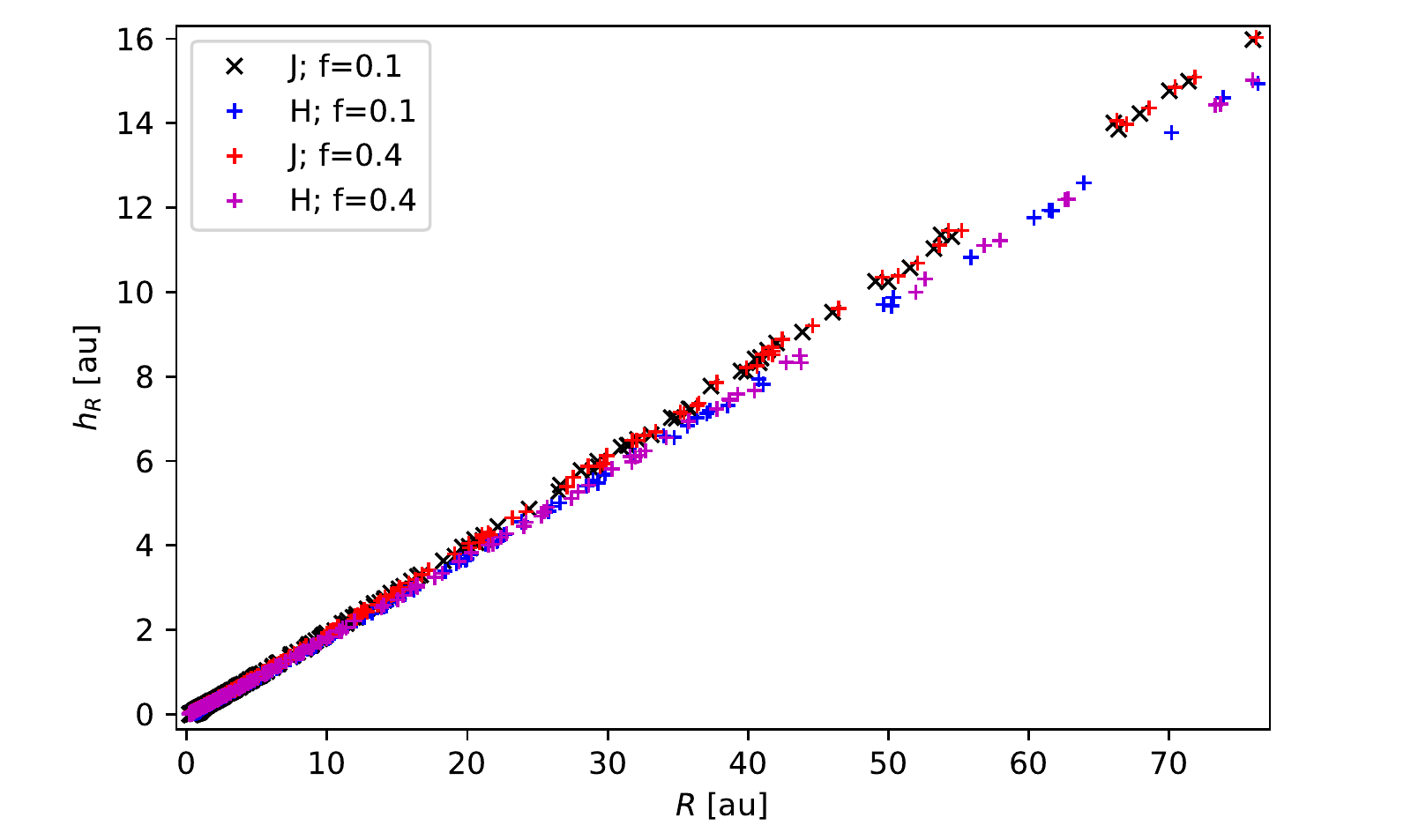}\\
    \includegraphics[trim=0.3cm 0.cm 1.2cm 1.0cm, clip=true,width=0.48\textwidth]{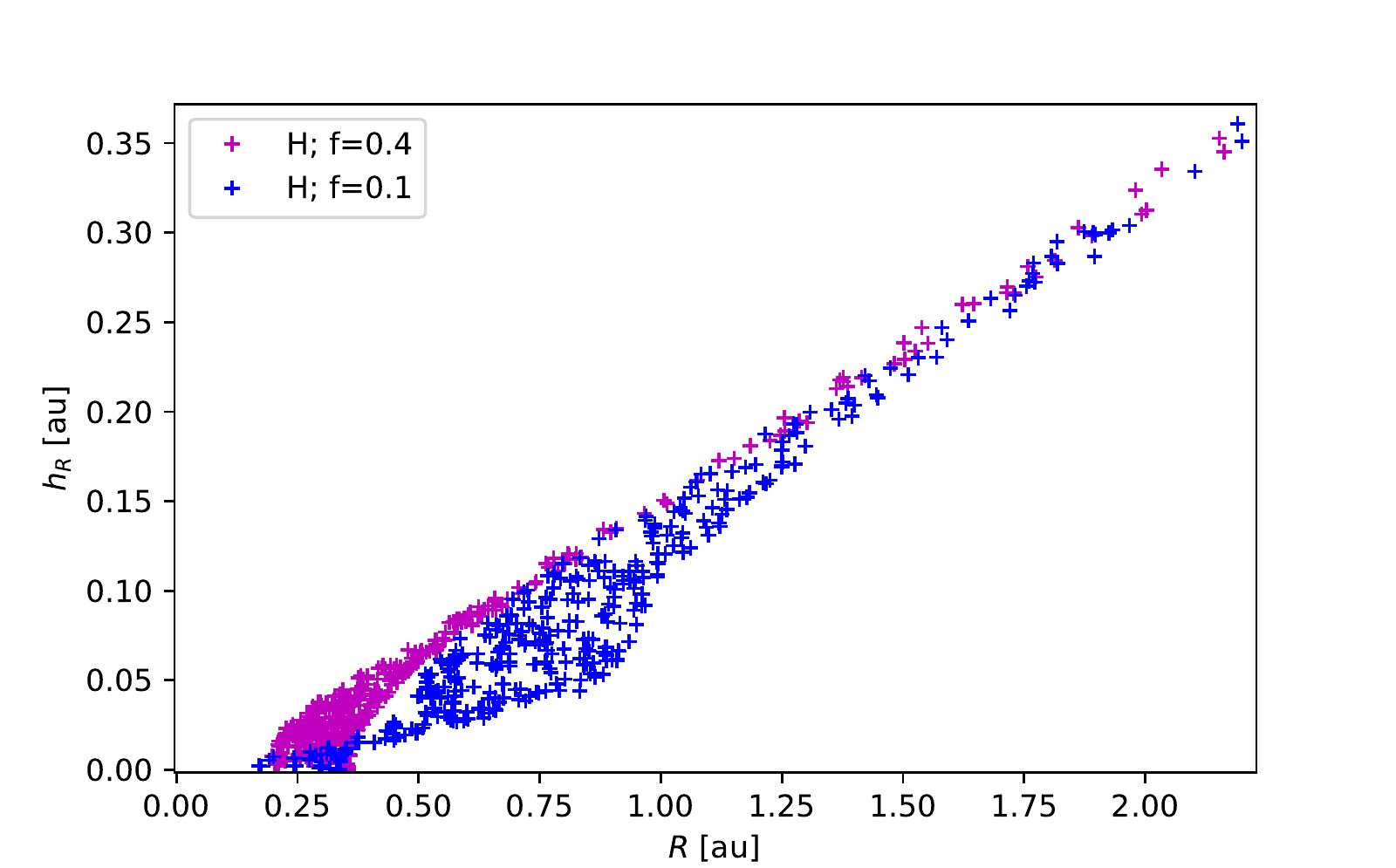}
    \caption{Top: comparison between the $J$- and $H$-band scattering surfaces resulting from TORUS models with different values of $f$. All other model parameters were identical in the two models. Bottom: zoom-in on the inner $\sim2\,$au of the $H$-band scattering surface traced by our ray-tracing algorithm. The horizontal spread in data points highlights the partially optically thick nature of the upper layers of the dust sublimation rim.}
    \label{fig:settleScatHeight}
\end{figure}

The $Q_{\phi}$ and $U_{\phi}$ images are unaffected by changes in $f$. We used a ray-tracing algorithm in TORUS to trace the $\tau_{\nu}=1.0$ scattering surface at $J$- and $H$-band and found that the height of the scattering surface remained unchanged beyond $\approx1.5-2\,$au (Figure\ref{fig:settleScatHeight}) when increasing the value of $f$ from $0.1\,h_{0, \rm{gas}}$ to $0.4\,h_{0, \rm{gas}}$. Differences between these models are observed in the SED (shortward of $\sim20\,\mu$m) and the visibilities. As the value of $f$ is increased, the total NIR to mid-IR flux in the SED drops. For $f\gtrsim0.3$, the inner disc rim broadly resembles the curved rim in the models of e.g. \citet{Isella05}, \citet{Tannirkulam07}, and \citet[][see the surface traced by the magenta data points in Figure~\ref{fig:settleScatHeight}]{Kama09}. For smaller values of $f$, the rim has a ``stepped'' feature (see the surface traced by blue data points in Figure~\ref{fig:settleScatHeight}), more akin to the disc rim model of \citet{McClure13}. Here, a more tightly curved sublimation rim forms with inner edge at $r_{\rm{in}}$ and a more loosely curved surface emerges above this at $r>r_{\rm{in}}$. The emergence of the small grains out of the settled rim extends the mas-scale NIR brightness distribution to larger scales. This improves the fit to the shorter baseline visibilities which are over-resolved in the models using larger $f$ values. 

\subsection*{Maximum size of surface-layer dust grains}
\begin{figure}
    \centering
    \includegraphics[trim=0.0cm 0.0cm 0.0cm 0.0cm, clip=true,width=0.35\textwidth]{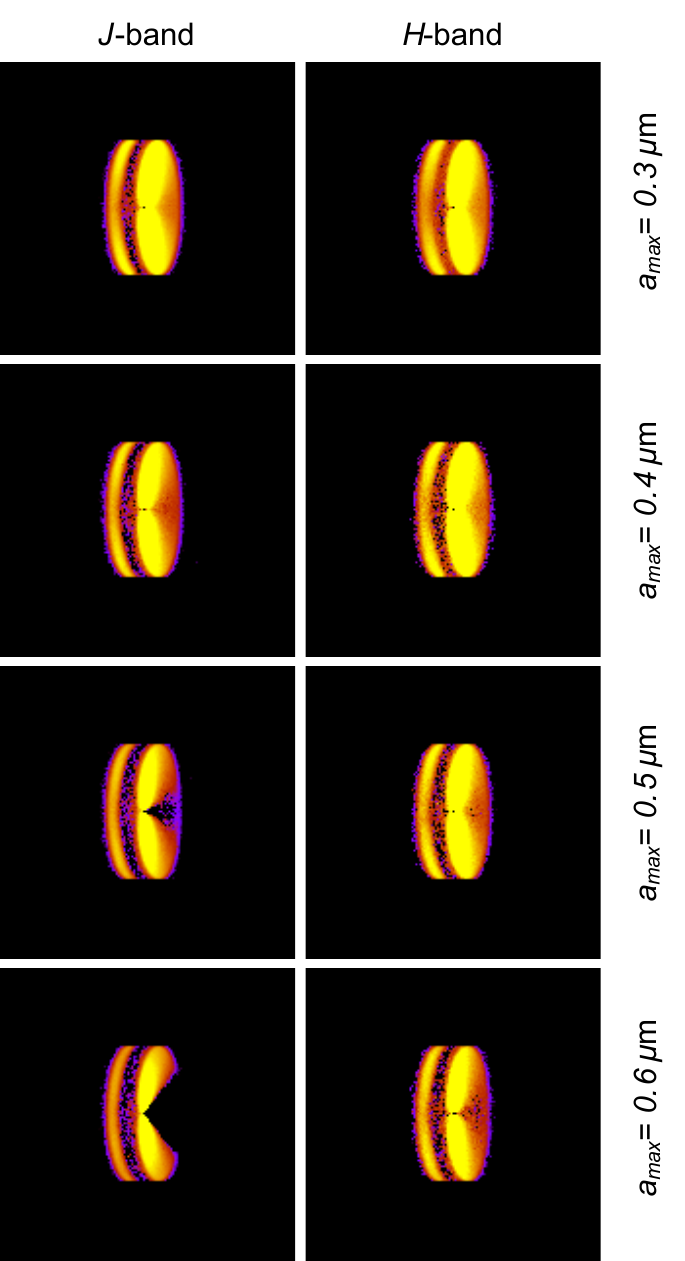}
    \caption{Impact of changing $a_{\rm{max}}$ on $J$- and $H$-band model $Q_{\phi}$ images (left and right panels, respectively). From top to bottom, the maximum grain size of non-settled grains is increased from $0.3\,\mu$m to $0.6\,\mu$m. Each image is $2\times2$''.}
    \label{fig:pacmanQphi}
\end{figure}

The value of $a_{\rm{max}}$ affects the radius at which the small grains emerge above the settled disc rim. \citet{Isella05} previously showed that larger grains can survive at higher temperatures as they are more efficient at cooling. This same process is responsible for the effect we see here. Increasing $a_{\rm{max}}$ therefore affects the NIR flux level and the shape of the visibilities in much the same way as seen for single grain size models \citep{Isella05, Davies18, Davies20b}. Changing $a_{\rm{max}}$ also affects the scattering phase function: at the relatively high inclinations we explore, we see an asymmetric $Q_{\phi}$ brightness distribution and increasing $a_{\rm{max}}$ results in a decrease in back-scattering efficiency, relative to forward scattering (Figure~\ref{fig:pacmanQphi}). The extent of this difference is consistently more marked in the $J$-band image than the $H$-band. This is associated with the dependence of the scattering phase function on the grain size \citep{Stolker16b, Tazaki19}. One can also see from Figure~\ref{fig:pacmanQphi} that the arc features on the eastern side of the $Q_{\phi}$ images are dimmer for models with larger $a_{\rm{max}}$. The $U_{\phi}$ images also decrease in brightness with increasing $a_{\rm{max}}$. 

\subsection{Best-fitting TORUS model}\label{sec:RTRes}
\begin{figure*}
    \centering
    \includegraphics[trim=0.0cm 0.0cm 0.0cm 0.0cm, clip=true,width=0.35\textwidth]{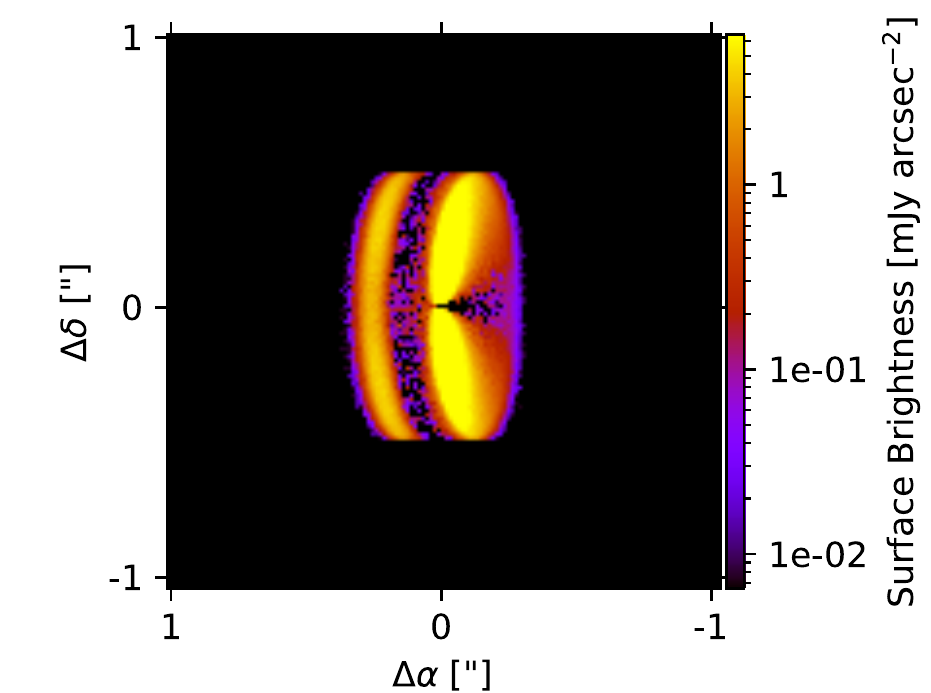}
    \includegraphics[trim=0.0cm 0.0cm 0.0cm 0.0cm, clip=true,width=0.35\textwidth]{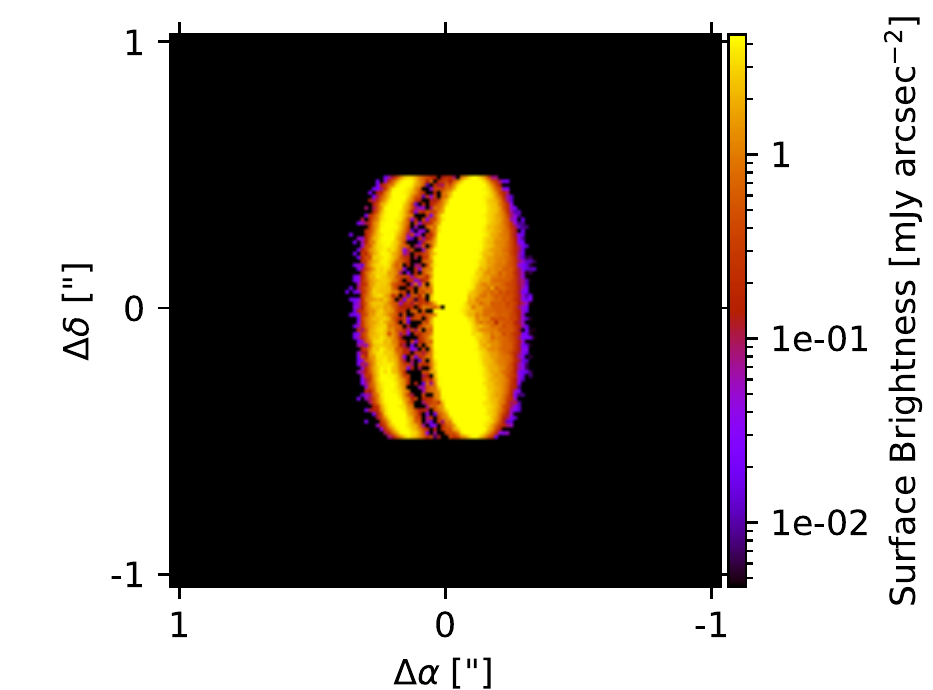}\\
    \includegraphics[trim=0.cm 0.cm 0.cm 0.cm, clip=true,width=0.35\textwidth]{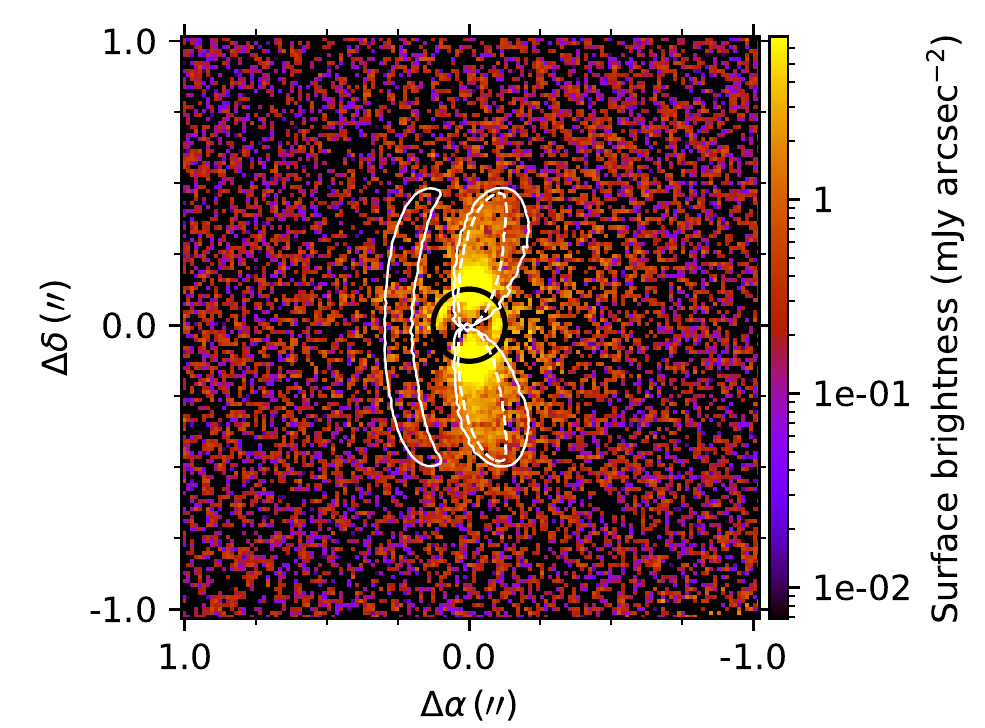}
    \includegraphics[trim=0.cm 0.cm 0.cm 0.cm, clip=true,width=0.35\textwidth]{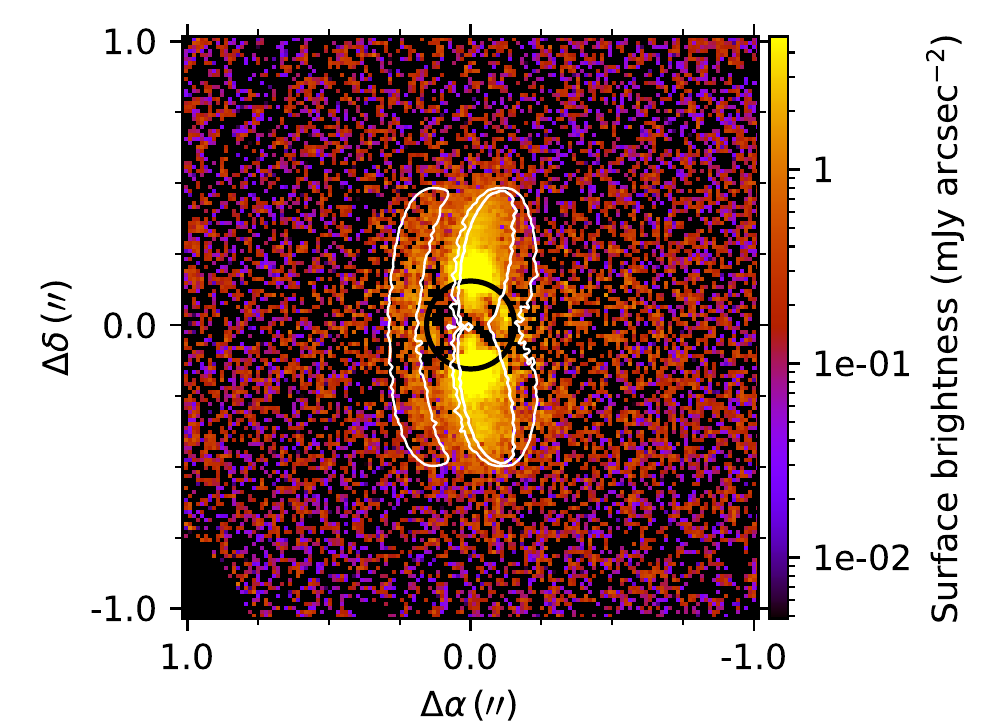}\\
    \includegraphics[trim=-0.3cm 0.cm 0.3cm 0.cm, clip=true,width=0.35\textwidth]{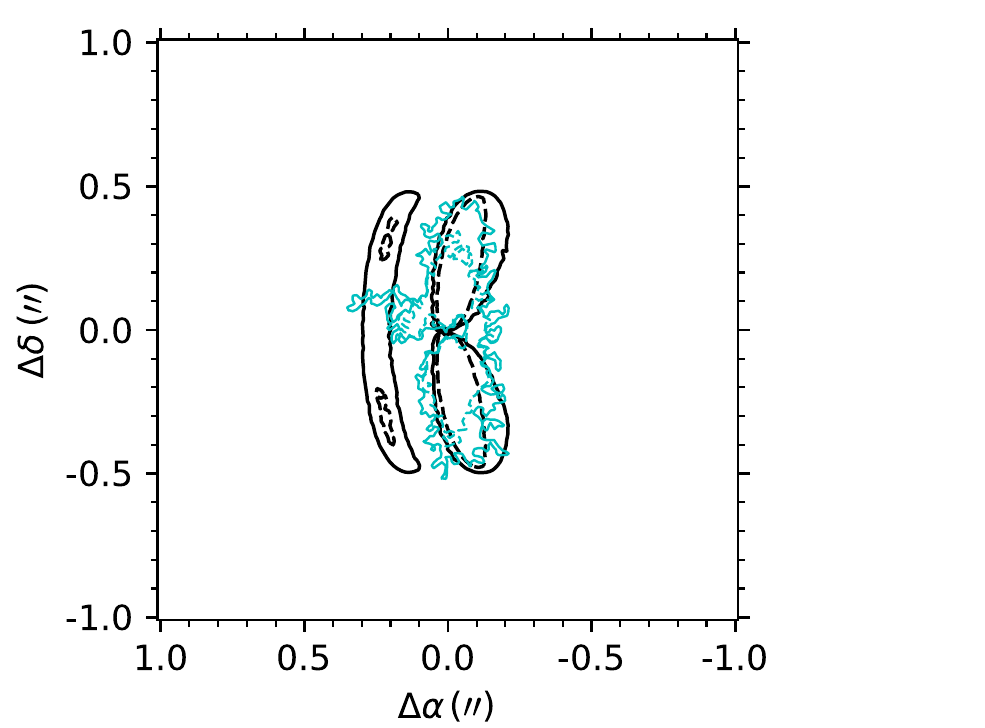}
    \includegraphics[trim=-0.3cm 0.cm 0.3cm 0.cm, clip=true,width=0.35\textwidth]{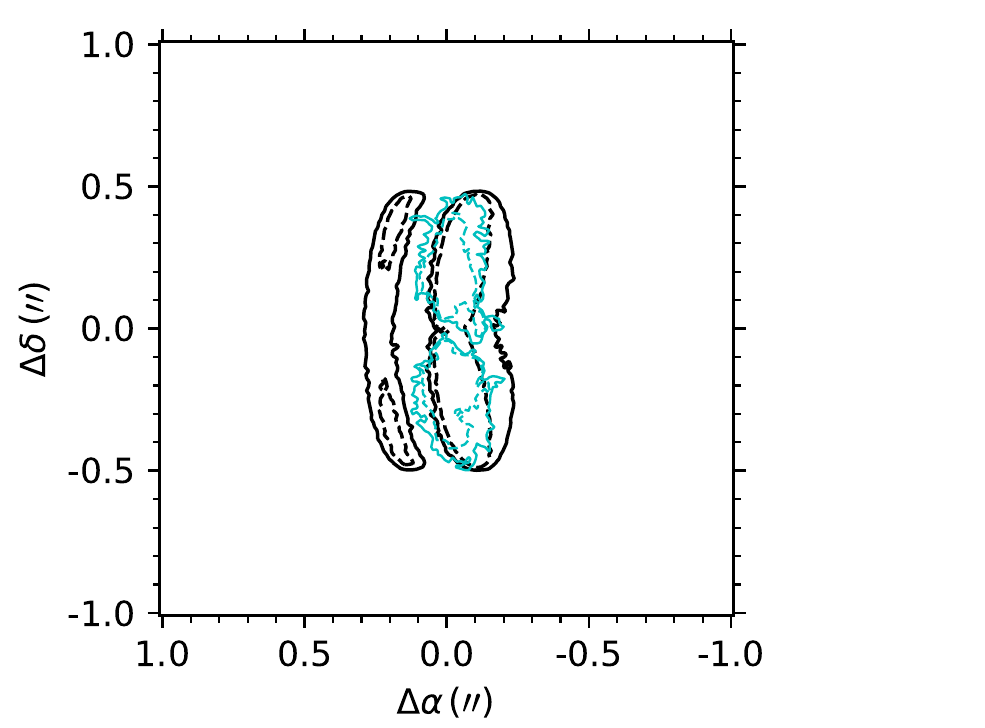}
    \caption{Left: $J$- (top) and $H$-band (bottom) $Q_{\phi}$ images for our best TORUS model ($h_{0, \rm{gas}}=10\,$au, $\beta=1.15$, $R_{\rm{out}}=75\,$au, $a_{\rm{max}}=0.50\,\mu$m, $i=72^{\circ}$ and PA$\,=0^{\circ}$). Middle: observed $Q_{\phi}$ images overlaid with $0.8$ and $4.0\,\rm{mJy}\,\rm{arcsec}^{-2}$ surface brightness contours extracted from these model images (white solid and dashed lines, respectively). Right: zoomed-in view comparing the $0.8$ and $4.0\,\rm{mJy}\,\rm{arcsec}^{-2}$ contours extracted from the model (black solid and dashed lines, respectively) and the $0.8$ and $1.8\,\rm{mJy}\,\rm{arcsec}^{-2}$ contours extracted from the observed $Q_{\phi}$ images (cyan solid and dashed lines, respectively). }
    \label{fig:RTbest_qphi}
\end{figure*}

\begin{figure}
    \centering
    \includegraphics[trim=0.0cm 0.0cm 0.0cm 0.0cm, clip=true,width=0.35\textwidth]{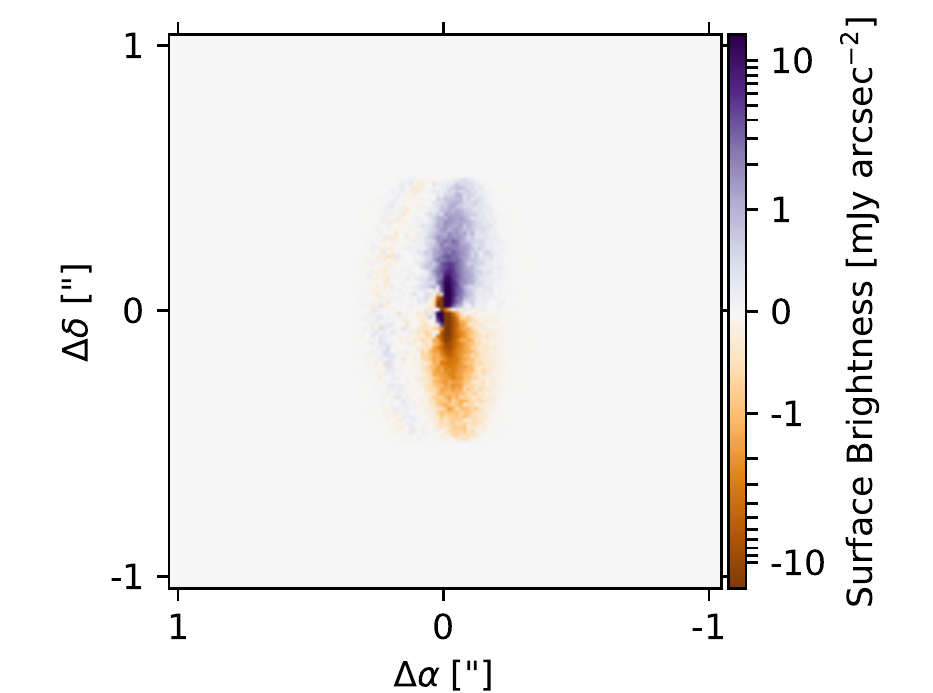}\\
    \includegraphics[trim=0.0cm 0.0cm 0.0cm 0.0cm, clip=true,width=0.35\textwidth]{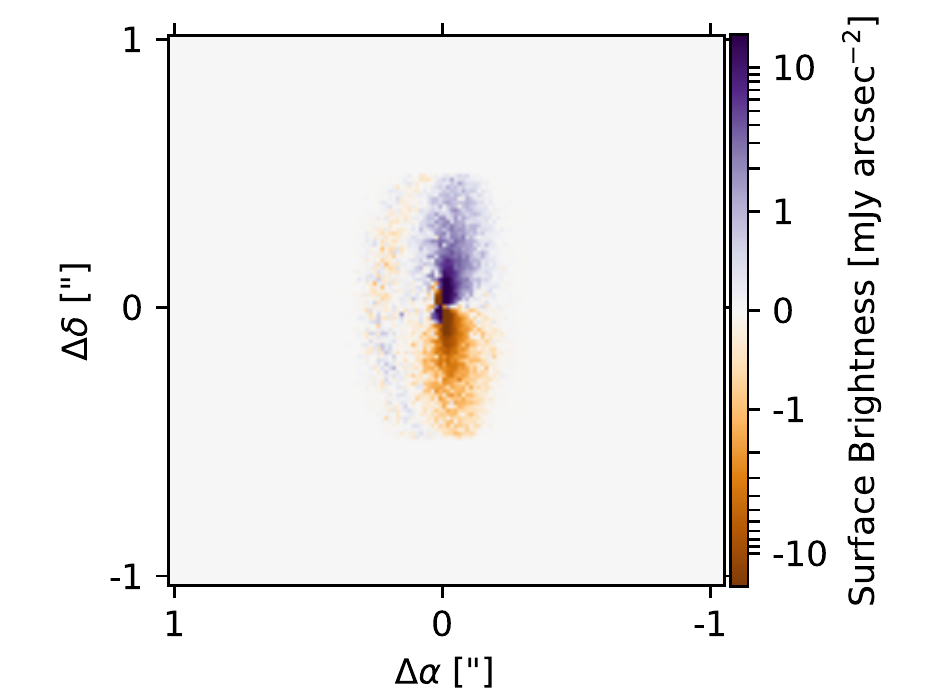}
    \caption{Model $U_{\phi}$ images for our best TORUS model ($h_{0, \rm{gas}}=10\,$au, $\beta=1.15$, $R_{\rm{out}}=75\,$au, $a_{\rm{max}}=0.50\,\mu$m, $i=72^{\circ}$ and PA$\,=0^{\circ}$). Top: $J$-band; bottom: $H$-band.}
    \label{fig:RTbest_uphi}
\end{figure}

\begin{figure}
    \centering
    \includegraphics[trim=0.3cm 0.0cm -0.2cm 0.2cm, clip=true,width=0.45\textwidth]{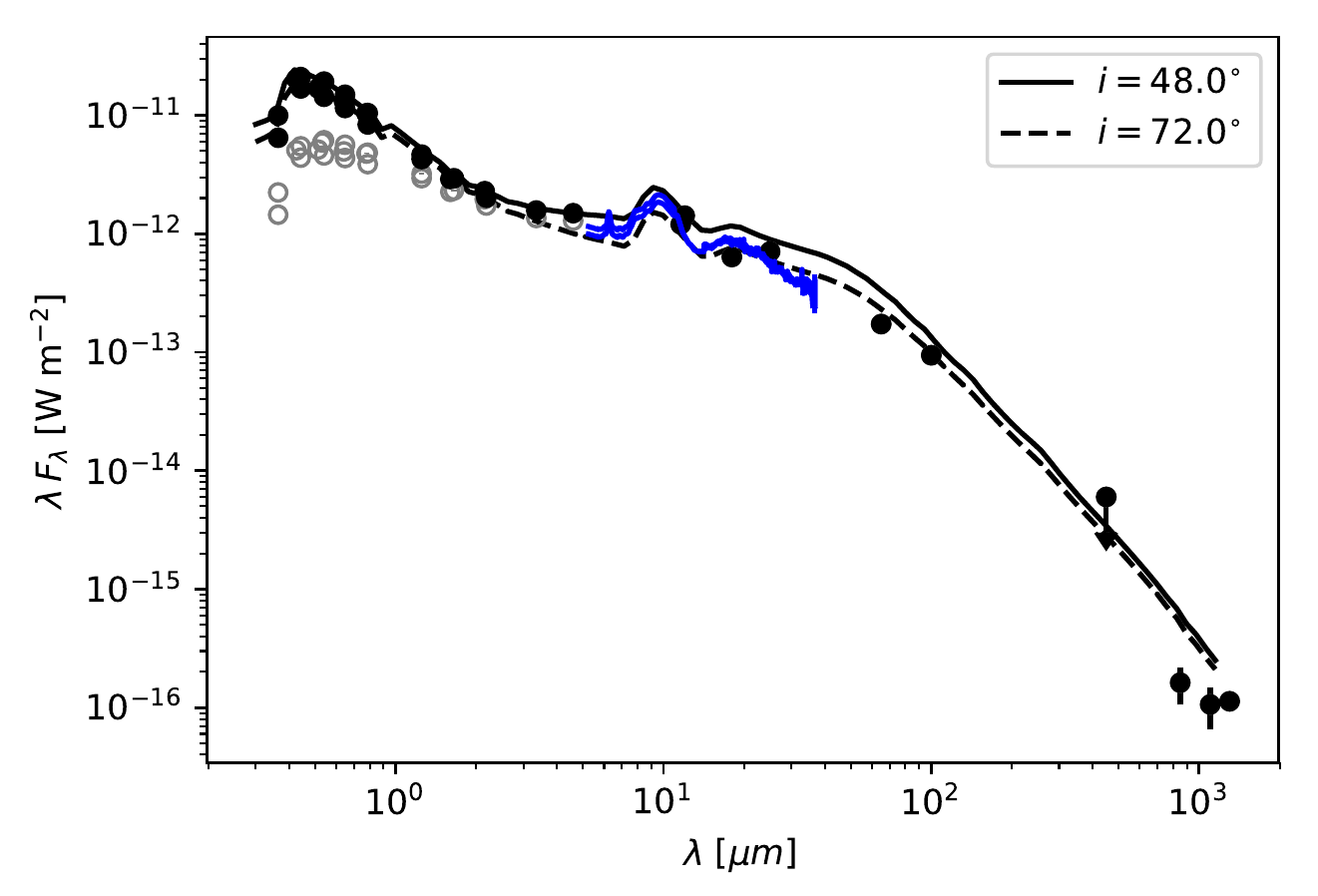}\\
    \includegraphics[width=0.45\textwidth]{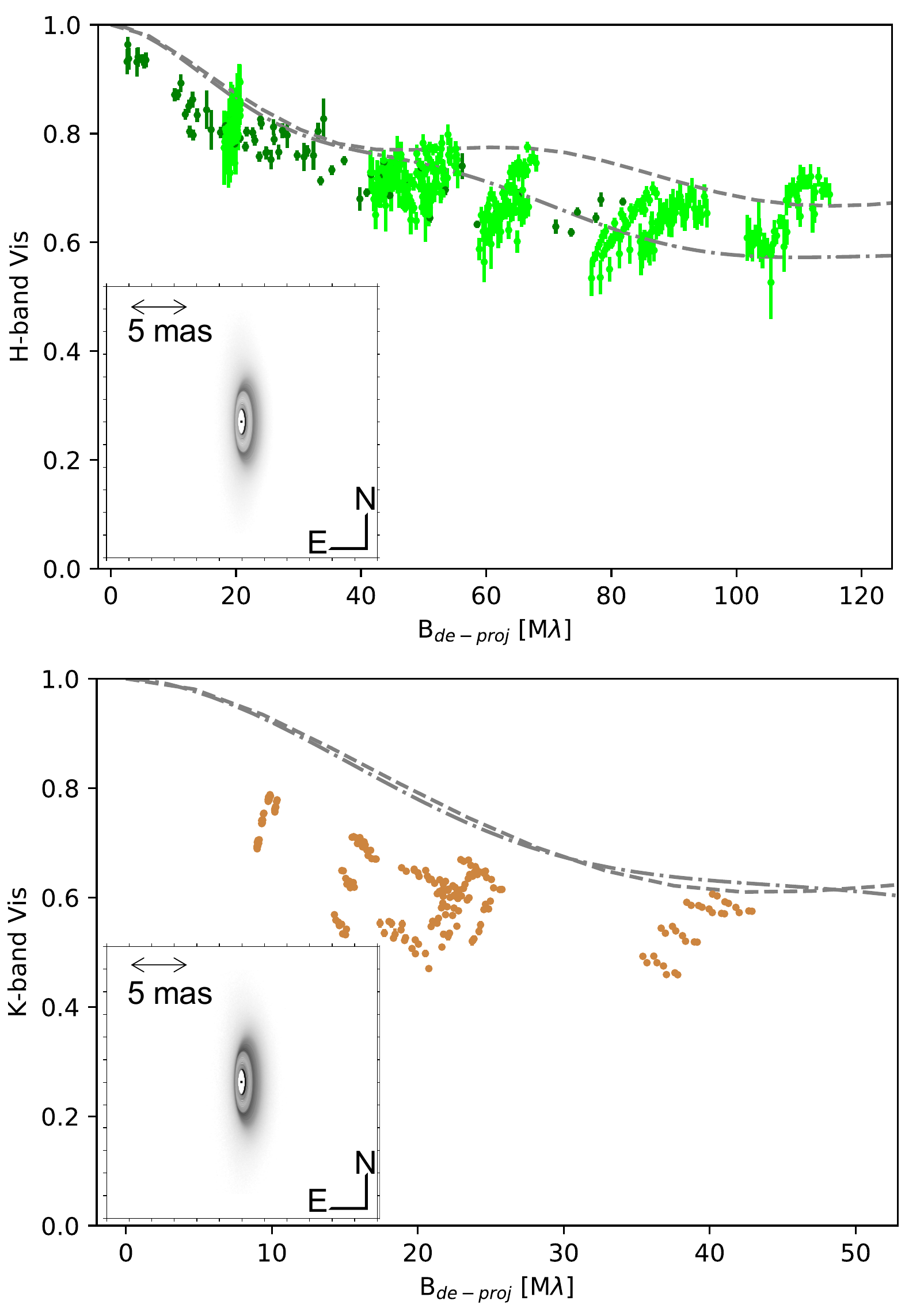}
    \caption{SED (top) and visibilities (middle and bottom) of our best TORUS model ($h_{0, \rm{gas}}=10\,$au, $\beta=1.15$, $R_{\rm{out}}=75\,$au, $a_{\rm{max}}=0.50\,\mu$m, $i=72^{\circ}$ and PA$\,=0^{\circ}$) compared to the observed spectro-photometry and NIR interferometry. Top: model SEDs at $i=72^{\circ}$ (dashed black line) and $i=48^{\circ}$ (solid black line) are compared to the \emph{Spitzer} spectrum (blue line) and photometry (open grey circles have zero de-reddening applied; black filled circles have been de-reddened using $A_{\rm{V}}=0.89$ and $R_{\rm{V}}=3.1$ - see Table~\ref{tab:inputParams}). Middle and bottom: deprojected visibility profiles extracted from the $1.65\,\mu$m (middle) and $2.13\,\mu$m (bottom) model total intensity images along baseline position angles tracing the major (grey dashed line) and minor (grey dot-dashed line) disc axes. Visibilities on the shortest baselines are overestimated by the model, indicating an additional extended emission component, unaccounted for in our models, is present. Colours represent the different beam combiners: see Figure~\ref{fig:uv}. The $24\times24\,$mas-scale model total intensity images are shown inset at the lower left corner of each plot.}
    \label{fig:RTbest_sedOLBI}
\end{figure}

The TORUS model providing the best overall fit was found to have $h_{0, \rm{gas}}=10\,$au, $\beta=1.15$, $R_{\rm{out}}=75\,$au, $a_{\rm{max}}=0.50\,\mu$m, $i=72^{\circ}$ and PA$\,=0^{\circ}$. The corresponding model $Q_{\phi}$ and $U_{\phi}$ images are compared to our GPI observations in Figures~\ref{fig:RTbest_qphi} and \ref{fig:RTbest_uphi}, respectively. The model SED and mas-scale images are compared to the respective observational data in Figure~\ref{fig:RTbest_sedOLBI}. 

The model $Q_{\phi}$ image is able to broadly replicate the $S_{\nu}$ level, location and extent of the main $E_{\rm{H}}$ and $A_{\rm{H}}$ features in the observed $H$-band image (Figure~\ref{fig:RTbest_qphi}). Reducing $i$ or increasing the height of the scattering surface by increasing $h_{0, \rm{gas}}$ and/or $\beta$ causes the separation between the elliptical and the arc feature in the model image to increase. Increasing $a_{\rm{max}}$ decreases the overall $S_{\nu}$ in the $Q_{\phi}$ and $U_{\phi}$ images and reduces the back-scattering efficiency, resulting in a larger dark portion on the west side of the elliptical feature in the $J$- and $H$-band model $Q_{\phi}$ images, like those seen in Figure~\ref{fig:pacmanQphi}. The weaker back-scattering we observe in the disc of HD~145718 has also been observed in scattered light imaging of other inclined discs (e.g. DoAr~25 \citealt{Garufi20}; IM~Lup \citealt{Avenhaus18}), suggesting that the surface layers of discs may be routinely populated by grains of size, $a\gtrsim\lambda/2\pi$. These larger grains possibly have an aggregate structure which provides them with aerodynamic support against settling. Alternatively, this may indicate a relative dearth of smaller grains ($a<<\lambda/2\pi$). For instance, \citet{Wolff21} found smaller dust grains were confined to a more diffuse region above the disc surface in their modelling of SSTC2D~J163131.2-242627. We do not see evidence of such a diffuse region around HD~145718.

None of the models we explored were able to reproduce the four quadrants of emission seen in our $U_{\phi}$ images. Instead, we note that the eastern quadrants of emission in the $U_{\phi}$ image in Figure~\ref{fig:gpi_obs} appear to overlap with the dark lane between the front and rear sides of the disc. As we discuss in Appendix~\ref{app:starpolsub}, the shape and $S_{\nu}$ levels in our $U_{\phi}$ are sensitive to our method of stellar and instrument polarisation correction. Without higher angular resolution observations from instruments with improved instrument polarisation characterisation, we are unable to assess whether this discrepancy is due to some underlying astrophysical process or an artifact of imperfect calibration.

The model SED is displayed as the black dashed line in the top panel of Figure~\ref{fig:RTbest_sedOLBI} while the photometric data are shown as black filled circles (de-reddening applied assuming $A_{\rm{V}}=0.89\,$mag with $R_{\rm{V}}=3.1$) or grey open circles (no de-reddening applied). Our model SED reproduces the shape of the full SED well. We also show the same model SED computed at a lower inclination of $48^{\circ}$ (solid black line). The slightly lower flux across optical wavelengths between the low and intermediate inclination model SEDs indicates that the circumstellar disc slightly occults the star along our line of sight in this model. Multi-colour photometric monitoring \citep[e.g.][]{Petrov19} and/or contemporaneous photometric and polarimetric/interferometric monitoring is required to confirm whether dust in the surface layers of disc obscures the star even during bright epochs. 

The visibility profiles in Figure~\ref{fig:RTbest_sedOLBI} (middle and bottom panels) are displayed as a function of deprojected baseline, 
\begin{equation}\label{eq:deprojB}
    B_{\rm{de-proj}} = B\left[ \sin^{2}(\phi)+\cos^{2}(i)\cos^{2}(\phi)\right]^{1/2}.
\end{equation}
Here, $B$ is the baseline length, $i$ is the disc inclination, and $\phi$ is the difference between the baseline position angle and the disc minor axis position angle. By displaying the visibility as a function of $B_{\rm{de-proj}}$, we account for the foreshortening of the brightness distribution along baseline position angles which trace the disc minor axis. Any vertical spread in visibility still present in the plot should then reflect a wavelength-dependence in the data, azimuthal variations in the flux contrast, imperfect calibration of the data, or noise. As our mas-scale model images are computed at a single wavelength ($1.65\,\mu$m for $H$-band; $2.13\,\mu$m for $K$-band), we implicitly assume a greybody approximation to the visibilities in each waveband. To account for the azimuthal variations in image brightness seen in our mas-scale model images, we extracted visibility curves along baseline position angles which trace the disc major and minor axes. These are shown as grey dashed and dot-dashed lines in the middle and bottom panels of Figure~\ref{fig:RTbest_sedOLBI}, respectively.

None of the models we explored could simultaneously reproduce both the sharp drop in visibility for $B_{\rm{de-proj}}\lesssim40\,\rm{M}\lambda$ and the relatively flat visibility profile beyond. However, models with $f<0.3\,h_{0,\rm{gas}}$ are able to reproduce the general shape of the visibilities at longer baselines and recover the visibility level on the longest $H$-band baselines well. The value of $f$ we infer from our best-fitting model ($0.1\,h_{0, \rm{gas}}$) is consistent with recent results from \citet{Villenave20} where the vertical extents of mm grains in a sample of edge-on discs were found to be on the order of a few au at $100\,$au.

Our inability to recover the sharp drop in visibility at short baselines suggests the presence of more extended circumstellar NIR emission than we are able to produce with our current models, as previously indicated by \citet{Lazareff2017pv} and \citet{Kluska20}. This may indicate the presence of a photoevaporative or magneto-hydrodynamically-driven disc wind, like that inferred for SU~Aur based on similar analyses \citep{Labdon19}. However, unlike for SU~Aur \citep{Ginski21}, we see no evidence of an outflow on larger scales in our GPI images but we note that the image regions along the disc minor axis are most affected by the choice of stellar polarisation subtraction (Appendix~\ref{app:starpolsub}). Alternatively, it may indicate that our non-settled, $0.01\geq a\geq 0.50\,\mu$m-sized dust grain mixture does not fully describe the disc surface, at least in the innermost disc regions. If we were to further segregate our dust prescription by grain size into three or more populations, we would likely see a smaller grain population emerge above this surface at larger radii, thus extending the NIR emitting region. Further investigation of this necessitates detailed theoretical work to simulate the combined effects of coagulation, settling, radial drift, collisional fragmentation and sublimation to predict where differently sized grains exist in the sublimation rim. 

\section{Discussion}\label{sec:discussion}
\subsection{Disc orientation and dust obscuration}\label{sec:geometry}
Our GPI images of HD~145718 reveal an inclined disc with its major axis oriented along a North--South direction. In Section~\ref{sec:analytical}, we fit elliptical ring models to coordinates tracing isophotes of surface brightness in the $J$- and $H$-band $Q_{\phi}$ images. Specifically, the full list of coordinates was trimmed to avoid features within (or close to the edge of) the inner working angle of the coronograph and the western portion of the $E_{\rm{J}}$ and $E_{\rm{H}}$ features, which deviated from an elliptical shape (Figure~\ref{fig:isophoteFit}). From our radiative transfer analysis in Section~\ref{sec:RTmodeling}, we saw that this deviation is attributable to the scattering phase function of the dust grains in the surface layers of the disc (Figure~\ref{fig:pacmanQphi}). The strong forward scattering and weak back-scattering we observe is typical of grain mixtures dominated by grains of size $a\gtrsim \lambda/2\pi$. These large grains are likely porous, providing them with aerodynamic support against settling \citep[see e.g.][]{Mulders13}. 

We infer a disc inclination in the range $67-71^{\circ}$, with major axis position angle between $1.0^{\circ}$ west of north and $0.6^{\circ}$ east of north. These results are consistent with previous assessments based on mm continuum and $K$-band interferometry \citep{Gravity19, Ansdell20}. Why \citet{Lazareff2017pv} and \citet{Kluska20} infer lower inclinations from $H$-band VLTI/PIONIER data is unclear: the NIR continuum emitting regions at $H$- and $K$-band are expected to be roughly coincident and therefore strongly aligned. If the lower inclination of $48^{\circ}$ measured by \citet{Kluska20} is used to de-project the baseline (Equation~(\ref{eq:deprojB})), the PIONIER and MIRC-X are not observed to follow the approximately Gaussian profile we see in the middle panel of Figure~\ref{fig:RTbest_sedOLBI} when we use $i=72^{\circ}$, regardless of the PA. We note that \citet{Kluska20} estimates the disc geometry from reconstructed images while \citet{Gravity19} and \citet{Ansdell20} perform their analysis in the Fourier plane. To reliably recover the inclination of highly inclined discs when using image reconstructions, one must ensure that the minor axis is well-resolved. Otherwise, as in this case, the emission along the minor axis is smoothed out by the interferometric beam and the disc will appear less inclined. 

At the inclination we infer, the surface layers of the inner disc of HD~145718 partially obscure the star along the observer's line of sight. This is most clearly seen in the top panel of Figure~\ref{fig:RTbest_sedOLBI} where the low inclination ($i=48^{\circ}$) model SED (black solid line) is slightly higher than the $i=72^{\circ}$ model (black dashed line) across optical wavelengths. This difference is within the allowed range of the de-reddened multi-epoch photometry (filled data points). In estimating the extinction, stellar luminosity and radius, we had found that differences between the bright and faint epoch photometry could be explained by a difference in the value of the total-to-selective extinction (Appendix~\ref{app:starparam}). Similarly, HD~145718 is known to exhibit UX~Ori and dipper variability, typically attributed to aperiodic stellar occultation by circumstellar dust close to the star \citep[e.g.][]{Dullemond03, Tambovtseva08}. What the data considered herein are not able to definitively assess is whether the stellar surface is always at least partially obscured by circumstellar material. Multi-colour photometric monitoring of HD~145718 such as that undertaken for RY~Tau and SU~Aur \citep{Petrov19} would be useful to establish this. 

\subsection{Assessing the robustness of our isophote fitting procedure}\label{subsec:surface}
\begin{figure}
    \centering
    \includegraphics[trim=0.8cm 1.6cm 1.2cm 2.5cm, clip=true,width=0.48\textwidth]{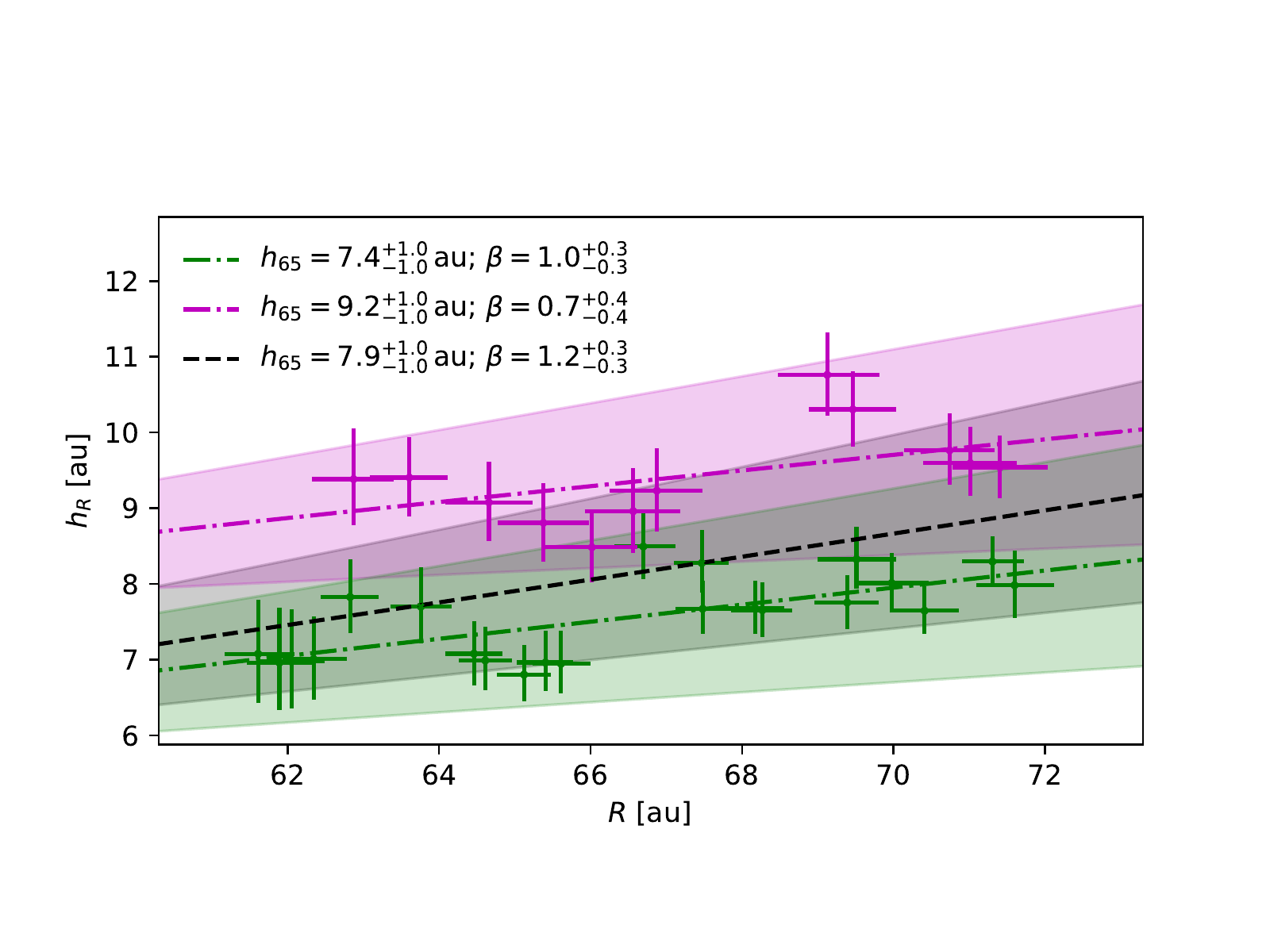}
    \caption{Apparent height of the scattering surface, $h_{\rm{scat}}(r)$, as a function of the elliptical ring radius, $r$ (data are from Table~\ref{tab:qphiScaleHeight}: magenta and green data points are from fits to the $J$-band and $H$-band isophotes, respectively). The dashed lines show the best-fit power law profiles (Equation~(\ref{eq:scaleheight})) to the $H$-band (green line), the $J$-band (magenta line), and the combined $J$- and $H$-band data (black line). The amplitude, $h_{65}$, and exponent, $\beta$, of the power law fits are provided in the legend.}
    \label{fig:flaring}
\end{figure}

In addition to assessing the geometry of the disc, our isophote fitting procedure (Section~\ref{sec:analytical}) allowed us to explore the vertical extent of the $J$- and $H$-band scattering surfaces in a much more time and computationally efficient manner than our Monte Carlo radiative transfer modelling allows. Our results from these simple models suggested that (i) the $J$-band scattering surface is more extended than the $H$-band scattering surface and (ii) there appears to be a slight increase in $h_{\rm{scat}}(r)$ with $r$ in both wavebands. We show both of these effects more clearly in Figure~\ref{fig:flaring} by plotting our best-fit $h_{\rm{scat}}(r)$ values against the associated best-fit $r$ values for each $S_{\rm{\nu}}$ in Table~\ref{tab:qphiScaleHeight}: magenta and green data points represent the $J$- and $H$-band data, respectively. In this section, we provide some background on why probing the vertical structure of the disc is important, assess the degree of flaring inferred from our isophote fits, and examine the robustness of our isophote fitting by repeating the procedure on the synthetic observations computed from our best fit TORUS model.

The vertical extent of the gaseous component of circumstellar discs is expected to follow a flared profile (Equation~(\ref{eq:scaleheight})). For instance, based on theoretical predictions for centrally irradiated, steady-state accretion discs, the power law exponent on the gas scale height, $\beta_{\rm{gas}}$, is expected to lie in the range $1.125$ to $1.3$ \citep[e.g.][]{Kenyon87, Chiang97}. Indeed, we find that the value of $\beta$ in our best-fit TORUS model ($=1.15$) lies in this range. However, even in scenarios where dust grains are well-coupled to the gas, these values of $\beta$ are not expected to also describe the flaring exponent of the scattering surface \citep[see e.g.][]{Avenhaus18}. This is because the gas pressure scale height depends only on the gas temperature, while the height of the scattering surface depends on the dust properties (e.g. opacity, scattering phase function etc). Thus, probing the vertical disc structure allows us to probe the properties of the dust content of the disc. In recent years, observational constraints on the degree of flaring in the surface layers of dusty and gaseous components of protoplanetary discs have begun to be made \citep{Ginski16, Avenhaus18, Pinte18, Villenave20, Rich21}. Elliptical models, similar to those we employ in Section~\ref{sec:analytical}, have been used to determine the height of the scattering surface in concentric ring features of discs exhibiting substructure, with power law profiles (Equation~(\ref{eq:scaleheight})) then used to constrain the flaring exponent of the scattering surface, $\beta_{\rm{scat}}$. At the resolution of current observations, HD~145718 does not show indications of disc substructure and so these same methods are not applicable.

Assuming, first of all, that our surface brightness isophotes do trace concentric disc annuli, we use the results of our isophote fitting in Table~\ref{tab:qphiScaleHeight} to derive an initial estimate for the flaring of the $J$- and $H$-band scattering surfaces close to the apparent outer edge of the disc. Specifically, we employ the power law parameterisation for the vertical height of the scattering surface (e.g. Equation~(\ref{eq:scaleheight})). Based on the range of $r$ values in Table~\ref{tab:qphiScaleHeight}, we adopt a canonical radius, $r_0=65\,$au such that $h_0$ is the height at $65\,$au (which we denote $h_{65}$). To estimate $\beta_{\rm{scat}}$ and $h_{65}$, we performed least-squares fitting to the linear relation
\begin{equation}
    \log_{10}\left(h_{\rm{scat}}(r)\right) = \log_{10}(h_{65}) + \beta_{\rm{scat}} \log_{10}(r/65\,\rm{au}).
\end{equation}
Specifically, we drew 10,000 realisations of $h_{\rm{scat}}(r)$ and $r$ from split-normal distributions, based on their lower and upper bounded errors, and repeated the fit each time. The best-fit values and uncertainties on $\beta_{\rm{scat}}$ and $h_{65}$ are then the median and $1\sigma$ upper and lower quartiles from these fits. Fitting the data in this manner allowed us to account for the asymmetric errors on $h_{\rm{scat}}(r)$ and $r$. The resulting profiles are shown by the dashed lines in Figure~\ref{fig:flaring} while the shaded regions illustrate the uncertainties on the best-fitting values: $h_{65}=9.2^{+1.0}_{-1.0}$ with $\beta_{\rm{scat}}=0.7^{+0.4}_{-0.4}$ for the $J$-band data (magenta dot-dashed line and shaded region); $h_{65}=7.4^{+1.0}_{-1.0}$ with $\beta_{\rm{scat}}=1.0^{+0.3}_{-0.3}$ for the $H$-band data (green dot-dashed line and shaded region); and $h_{65}=7.4^{+1.0}_{-1.0}$ with $\beta_{\rm{scat}}=1.0^{+0.3}_{-0.3}$ for the combined $J$- and $H$-band data (black dashed line and grey shaded region). These values of $\beta_{\rm{scat}}$ are within the range of previous observational results determined by fitting power laws to the heights of concentric ring features in discs with substructure: \citet{Avenhaus18} found $\beta_{\rm{scat}}=1.605\pm0.132$ for V4046~Sgr, $\beta_{\rm{scat}}=1.116\pm0.095$ for RX~J~1615.3–3255, and $\beta_{\rm{scat}}=1.271\pm0.197$ (IM~Lup) while \citet{Ginski16} found $\beta_{\rm{scat}}=1.73$ for CU~Cha.

\begin{figure}
    \centering
    \includegraphics[trim=0.5cm 0.cm 1.0cm 0.8cm, clip=true,width=0.48\textwidth]{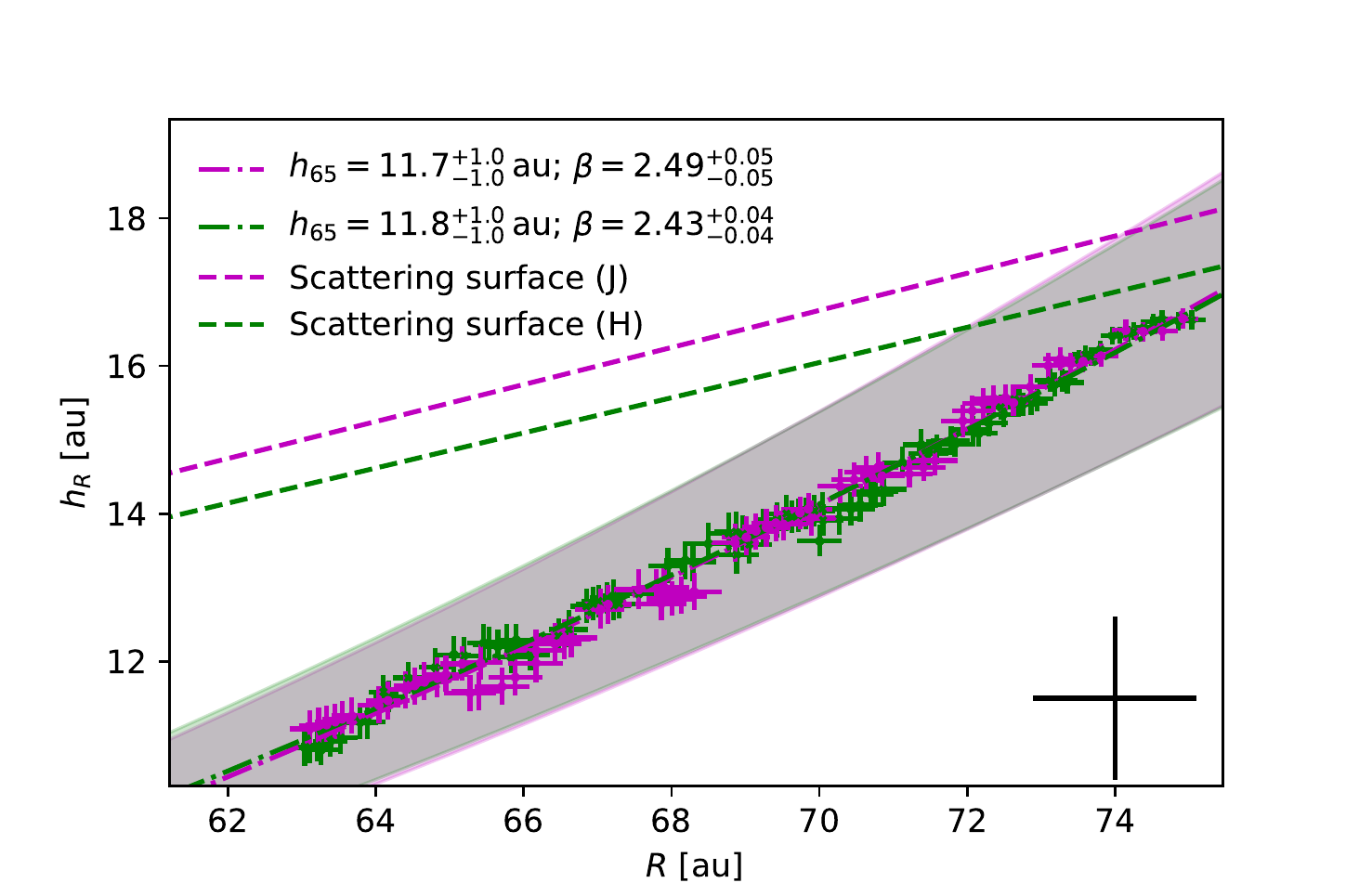}
    \caption{Application of our isophote fitting procedure (Section~\ref{sec:geometry}) to the synthetic $J$- and $H$-band $Q_{\phi}$ images computed from our best-fitting TORUS model. Radii and heights extracted from the $J$- and $H$-band images are plotted in magenta and green, respectively. The power law fits to these data are shown by the dot-dash lines, with the shaded areas representing the $1\sigma$ uncertainties on $h_{65}$ and $\beta_{\rm{scat}}$ (see plot legend). The $\tau_{\rm{scat}}=1$ scattering surfaces at $1.25\,\mu$m ($J$-band) and $1.65\,\mu$m ($H$-band), as measured by TORUS, are shown by the dashed lines. The cross-hair in the lower right corner represents the pixel scale in our GPI and TORUS images.}
    \label{fig:modQphiIso}
\end{figure}

We used our synthetic $Q_{\phi}$ TORUS images to examine whether our isophote fitting procedure is indeed tracing the height of the scattering surface over multiple disc radii (and therefore the flaring of the disc scattering surface). Using our synthetic $Q_{\rm{\phi}}$ images (top row of Figure~\ref{fig:RTbest_qphi}), we repeat the procedure outlined in Section~\ref{sec:analytical} to extract the coordinates of $S_{\nu}$ isophotes. We trimmed the data using the same constraints as before so as to reduce the effect of the low back-scattering efficiency. We kept $i$ and PA as free parameters and plot the resultant $r$ and $h_{\rm{scat}}(r)$ values in Figure~\ref{fig:modQphiIso} (we use magenta and green crosses to signify $J$- and $H$-band data, respectively). Again, we use Equation~(\ref{eq:scaleheight}), to estimate $h_{65}$ and $\beta_{\rm{scat}}$ from the results of the isophote fits. We find that the inferred $J$- and $H$-band scattering surfaces are coincident and the apparent flaring is much more pronounced, with $\beta\sim2.4-2.5$, casting doubt on the applicability of our isophote fitting procedure.

To further inspect our isophote modelling procedure, we  used a ray-tracing algorithm to compute the $\tau_{\rm{scat}}=1.0$ scattering surface at $1.25$ and $1.65\,\mu$m. We compare this to the apparent surface traced by our isophote fits to the best-fit synthetic $Q_{\rm{\phi}}$ images in Figure~\ref{fig:modQphiIso}. The $\tau_{\rm{scat}}=1.0$ scattering surfaces at $1.25$ and $1.65\,\mu$m are indicated by the dashed magenta and green lines, respectively. Two things are immediately clear: (i) our isophote fitting procedure does not recover the degree of flaring in the $\tau_{\rm{scat}}=1.0$ scattering surface; and (ii) the $J$-band scattering surface does extend to larger scale heights than the $H$-band scattering surface. The $J$-band surface has an aspect ratio of $\sim0.24$ while that of the $H$-band is $\sim0.22$. Our best-fit TORUS model has a gas scale height of $10\,$au at $100\,$au and, with $\beta_{\rm{gas}}=1.15$, this corresponds to a height of $\sim6.1\,$au at $65\,$au, indicating that the NIR scattering surface lies at a height of $\approx2.5$ gas pressure scale heights. 

Close to the outer edge of the disc, our isophote fitting method does provide a reasonable estimate of the vertical extent of the scattering surface, given the pixel resolution (indicated by the cross-hair in the lower right corner of Figure~\ref{fig:modQphiIso}) and the centering accuracy of the GPI coronograph (equivalent to $\sim0.47\,$au at a distance of $152.5\,$pc; see Section~\ref{sec:observations}). 

\begin{figure*}
    \centering
    \includegraphics[trim=0.0cm 0.1cm 0.0cm 0.0cm, clip=true,width=0.85\textwidth]{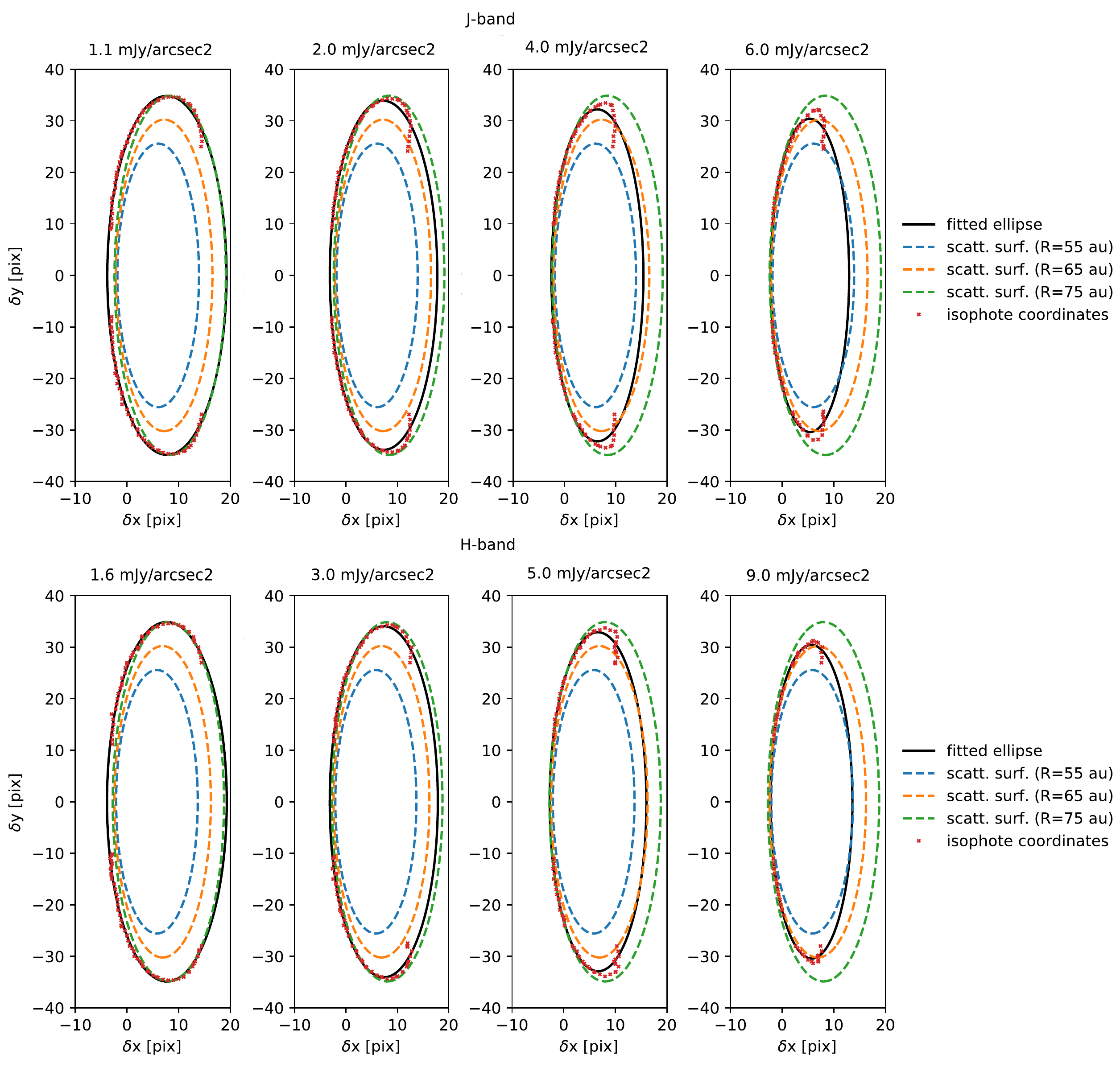}
    \caption{Coordinates of $S_{\nu}$ isophotes extracted from synthetic TORUS model $Q_{\phi}$ images (red crosses) and the resultant elliptical ring fits (solid black ellipse) are compared to ellipses traced by the wavelength-dependent $\tau_{\rm{scat}}=1.0$ scattering surface at $55\,$au (dashed blue ellipse), $65\,$au (dashed orange ellipse) and $75\,$au (dashed green ellipse). Top row: $J$-band; bottom row: $H$-band. The values of $S_{\nu}$ are provided above each subplot window.}
    \label{fig:examineIso}
\end{figure*}

To understand why our isophote fitting procedure does not recover the scattering surface traced by the ray-tracing algorithm, we determined the elliptical ring parameters of the ellipse drawn out by the $\tau_{\rm{scat}}=1.0$ scattering surface at $r=55$, $65$, and $75\,$au and compared these to the isophotes extracted from the model image at the same disc radii. These are shown in Figure~\ref{fig:examineIso}, overlaid with the $S_{\nu}$ isophote coordinates and corresponding best-fit ellipse. 

\begin{figure}
    \centering
    \includegraphics[trim=0.2cm 0.1cm 1.5cm 0.8cm, clip=true,width=0.45\textwidth]{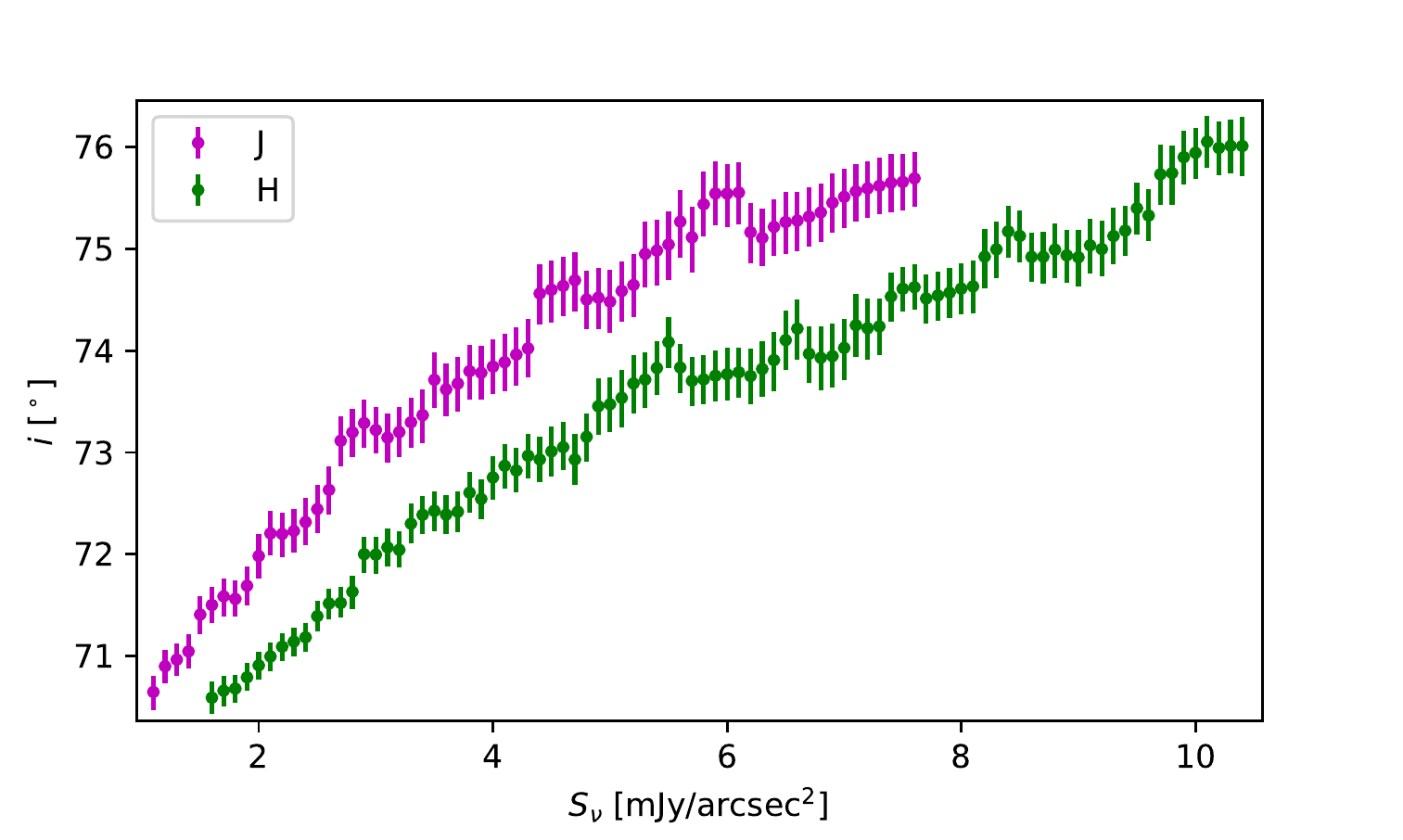}
    \caption{Trend observed between the surface brightness of the isophote, $S_{\nu}$, and the inferred inclination, $i$, when fitting elliptical ring models to the isophotes extracted from the synthetic $Q_{\phi}$ images computed using TORUS. }
    \label{fig:sb_v_inc}
\end{figure}

We find that for the smallest values of $S_{\nu}$ (left-most panel in Figure~\ref{fig:examineIso}), the coordinates tracing the eastern side of the ellipse appear offset further to the east compared to the $\tau_{\rm{scat}}=1.0$ scattering surface at $r=75\,$au. This is right at the outer edge of the disc ($R_{\rm{out}}=75\,$au; see Table~\ref{tab:RTparams}), indicating that these $S_{\nu}$ isophotes trace scattering events at the radial edge of the disc, rather than in the disc surface layers. This results in an overestimation of the width of the ellipse along its minor axis, therefore affecting the inferred values of $h_{\rm{scat}}(r)$ and $i$. Indeed, fits using larger values of $S_{\nu}$ tend to larger values of $i$ (see Figure~\ref{fig:sb_v_inc}). The second panel from the left in Figure~\ref{fig:examineIso} corresponds to the ellipse fit where the inferred inclination matches the prescribed value of $72^{\circ}$. Even here, the eastern portion of the ellipse is well traced but the isophote does not sufficiently trace the western portion of the apex of the ellipse, resulting in an underestimation of $h_{\rm{scat}}(r)$. The apparent truncation of the ellipse apex to the west is worse for larger $S_{\nu}$ isophotes, giving rise to the apparent $S_{\nu}$--$i$ relation we observe, and is attributable to the low back-scattering efficiency for the grain mixture used in this model (see Figure~\ref{fig:pacmanQphi}). The isophote fitting technique used in Section~\ref{sec:analytical} may therefore be improved by further requiring the ellipse apex to pass through the coordinates with the largest radial extent from the image centre and/or by independently constraining the disc inclination (e.g. using ALMA).

\section{Summary}\label{sec:summary}
We present a multi-instrument study of the intermediate mass YSO, HD~145718. Specifically, we complemented GPI $J$- and $H$-band polarised differential imaging from G-LIGHTS \citep{Monnier17, monnier2019, laws2020} with NIR interferometry (archival VLTI/GRAVITY and VLTI/PIONIER data plus new CHARA/MIRC-X observations), infrared spectroscopy and multi-band, multi-epoch photometry. 

Our GPI images reveal an inclined disc with major axis position angle close to $0^{\circ}$. The offset between the centroid of emission in the open face of the disc and the image centre indicates the disc scattering surface is elevated above the disc midplane. Further, the strong forward scattering and weak back-scattering evident in the $Q_{\phi}$ images indicates that large grains (size, $a\gtrsim \lambda/2\pi$) are abundant in the surface layers of the disc. 

Inspection of the multi-epoch $BVRI$ photometry indicated that previously published estimates of HD~145718's stellar luminosity and radius have been affected by direct line-of-sight occultation by circumstellar dust. Using the brightest epoch photometry retrieved by SEDBYS \citep{Davies21} and a total-to-selective extinction, $R_{\rm{V}}=3.1$, we re-estimated the visual extinction ($A_{\rm{V}}=0.89^{+0.34}_{-0.08}\,$mag), stellar radius ($R_{\star}=1.97^{+0.12}_{-0.11}\,\rm{R_{\odot}}$) and stellar luminosity ($L_{\star}=14.3^{+3.9}_{-3.1}\,\rm{L_{\odot}}$) for HD~145718. Furthermore, we find consistent estimates of $R_{\star}$ and $L_{\star}$ when applying the same method to faint epoch photometry from \citet{Vieira03} and using $R_{\rm{V}}=5.0$. This further supports the idea that the UX~Ori and dipper photometric variability displayed by HD~145718 is associated with obscuration by dust grains and that, during periods of occultation, the dust grains are larger, on average, than those in the interstellar medium \citep{Hernandez04}. Further multi-colour photometric monitoring, similar to that undertaken for RY~Tau and SU~Aur by \citet{Petrov19}, is required to determine whether smaller circumstellar grains still contribute to the extinction during the brightest epochs. 

We used an off-centre elliptical ring model to fit isophotes of surface brightness in the $Q_{\phi}$ images, finding:
\begin{itemize}
    \item the optically thick disc extends out to a radius of $\sim75\,$au, assuming a distance to HD~145718 of $152.5\,$pc \citep{Gaia18};
    \item the disc is oriented with major axis position angle between $-1.0^{\circ}$ and $0.6^{\circ}$ east of north and inclined at $67^{\circ}$ to $71^{\circ}$, consistent with previous measurements based on mm continuum and $K$-band interferometry \citep{Gravity19, Ansdell20};
\end{itemize}

We used detailed radiative transfer modelling to self-consistently investigate the radial and vertical disc structure and to assess the extent to which our isophote fitting could be used to probe the disc scattering surface. From our radiative transfer modelling, we found that a model comprising a centrally illuminated passive disc with gas pressure scale height, $h_{0, \rm{gas}}=10\,$au, flaring exponent, $\beta=1.15$, outer disc radius, $R_{\rm{out}}=75\,$au, maximum size of non-settled grains, $a_{\rm{max}}=0.50\,\mu$m, large-grain settling factor, $f=0.1\,\rm{h_{0}}$, inclination, $i=72^{\circ}$, and major axis position angle, PA$\,=0^{\circ}$ provides a good fit to:
\begin{itemize}
    \item the surface brightness, location and extent of the ellipse and arc features in the $J$- and $H$-band $Q_{\phi}$ images;
    \item the shape of the full SED from optical to millimetre wavelengths;
    \item the general shape and stellar-to-circumstellar flux contrast level traced by the $H$-band visibilities on the largest baselines probed;
    \item and the general shape of the $K$-band visibilities.
\end{itemize}

However, the model could not account for the immediate drop in visibility on the shortest baselines in both $H$- and $K$-bands. As previously suggested by e.g. \citet{Kluska20}, this likely indicates the presence of more extended NIR emission, potentially in the form of an outflow. We find no robust evidence of this outflow on the larger scales probed by our GPI images. Further assessment of this requires better assessment of the instrument polarisation which would allow us to improve our stellar polarisation subtraction.

By comparing synthetic images to the $Q_{\phi}$ images we obtained with GPI, we find that fitting ellipses to isophotes of $Q_{\phi}$ surface brightness recovers the general elliptical shape of the emission but, due to the azimuthally dependent scattering efficiency, cannot reliably recover both the disc inclination and scattering surface height. We propose that our simple isophote fitting method could be improved by independently constraining the disc inclination using e.g. ALMA continuum observations to counter the preference we observe for lower (higher) surface brightness isophotes to appear less (more) inclined. 

\section*{Acknowledgements}
The authors thank the anonymous referee for their comments which improved the content of this paper.
We thank Ren{\'e} Oudmaijer for useful discussions regarding the estimation of visual extinction and Fred C. Adams for a detailed reading of the manuscript. 
C.L.D acknowledges financial support from the College of Engineering, Mathematics and Physical Sciences at the University of Exeter. 
E.A.R and J.D.M acknowledge funding from a National Science Foundation grant (reference NSF-AST1830728). A.S.E.L acknowledges financial support from a Science Technology and Facilities Council (STFC) studentship (reference 1918673). 
J.B acknowledges support by NASA through the NASA Hubble Fellowship grant \#HST-HF2-51427.001-A awarded  by the Space Telescope Science Institute, which is operated by the Association of Universities for Research in Astronomy, Incorporated, under NASA contract NAS5-26555. 
This study is based on observations obtained at the Gemini Observatory, which is operated by the Association of Universities for Research in Astronomy, Inc., under a cooperative agreement with the NSF on behalf of the Gemini partnership: the National Science Foundation (United States), the National Research Council (Canada), CONICYT (Chile), the Australian Research Council (Australia), Ministério Ciência, Tecnologia e Inovação (Brazil) and Ministerio de Ciencia, Tecnología e Innovación Productiva (Argentina).
This work is based upon observations obtained with the Georgia State University Center for High Angular Resolution Astronomy Array at Mount Wilson Observatory.  The CHARA Array is supported by the National Science Foundation under Grant No. AST-1636624 and AST-1715788.  Institutional support has been provided from the GSU College of Arts and Sciences and the GSU Office of the Vice President for Research and Economic Development. MIRC-X received funding from the European Research Council (ERC) under the European Union's Horizon 2020 research and innovation programme (Grant No. 639889).
The calculations for this paper were performed on the University of Exeter Supercomputer, a DiRAC Facility jointly funded by STFC, the Large Facilities Capital Fund of BIS, and the University of Exeter. 
This work uses data obtained from the ESO Science Archive Facility.  
This research has made use of the NASA/IPAC Infrared Science Archive, which is funded by the National Aeronautics and Space Administration and operated by the California Institute of Technology;
the Jean-Marie Mariotti Center \texttt{OiDB} service\footnote{Available at http://oidb.jmmc.fr }; the SIMBAD database, operated at CDS, Strasbourg, France; 
the VizieR catalogue access tool, CDS, Strasbourg, France; 
NASA's Astrophysics Data System Bibliographic Services.


\section*{Data Availability}
The GPI data are available from the Gemini Observatory Archive at \url{https://archive.gemini.edu/searchform} and can be accessed using proposal number GS-2018A-LP-12. The MIRC-X data will be made available through the OiDB (\url{http://oidb.jmmc.fr}) following publication. VLTI/GRAVITY and VLTI/PIONIER data are available in the ESO archive (\url{http://archive.eso.org/cms.html}) and the OiDB and can be accessed using the target name `HD~145718'. The photometry are accessible through SEDBYS, available at \url{https://gitlab.com/clairedavies/sedbys}. The IR spectra are available in the IRSA (\url{https://irsa.ipac.caltech.edu}). 




\bibliographystyle{mnras}
\bibliography{hd145718} 



\appendix
\section{Photometry}\label{appen:phot}
The full list of photometry returned by SEDBYS is provided in Table~\ref{tab:phot}.

\begin{table}
 \centering
 \caption{Collated photometry. Individual references are provided in column 3.} \label{tab:phot}
 \begin{tabular}{lll}
  \hline
  $\lambda$ & $\lambda F_{\lambda}$   & Reference\\
  ($\mu$m)  & ($10^{-13}$\,W\,m$^{-2}$) &      \\
  (1) & (2) & (3) \\
  \hline
0.36 & $14$ & \citet{Vieira03} \\ 
0.36 & $22$ & \citet{Lazareff2017pv} \\ 
0.43 & $51\pm1$ &  \citet{Hog2000hg} \\ 
0.44 & $44$ & \citet{Vieira03} \\ 
0.44 & $55$ & \citet{Lazareff2017pv} \\ 
0.51 & $51.3\pm0.7$ & \citet{Gaia18} \\ 
0.53 & $58.8\pm0.9$ & \citet{Hog2000hg} \\ 
0.54 & $61$ & \citet{Lazareff2017pv} \\ 
0.54 & $46$ & \citet{Vieira03} \\ 
0.64 & $49.5\pm0.3$ & \citet{Gaia18} \\ 
0.65 & $56$ & \citet{Lazareff2017pv} \\ 
0.65 & $43$ & \citet{Vieira03} \\ 
0.78 & $47.4\pm0.5$ & \citet{Gaia18} \\ 
0.79 & $39$ & \citet{Vieira03} \\ 
0.79 & $48$ & \citet{Lazareff2017pv} \\ 
1.25 & $29$ & \citet{Lazareff2017pv} \\ 
1.25 & $32.1\pm0.7$ & \citet{Cutri2003ya} \\ 
1.60 & $22.5$ & \citet{Lazareff2017pv} \\ 
1.65 & $23.1\pm0.6$ & \citet{Cutri2003ya} \\ 
2.15 & $19.6\pm0.4$ & \citet{Cutri2003ya} \\ 
2.18 & $17.3$ & \citet{Lazareff2017pv} \\ 
3.35 & $13.6\pm0.7$ & \citet{Cutri2012ww} \\ 
4.60 & $12.9\pm0.5$ & \citet{Cutri2012ww} \\ 
9.00 & $13.88\pm0.04$ & \citet{Ishihara2010fp} \\ 
11.60 & $11.99\pm0.09$ & \citet{Cutri2012ww} \\ 
12.00 & $14.2$ & \citet{Oudmaijer92} \\ 
18.00 & $6.35\pm0.08$ & \citet{Ishihara2010fp} \\ 
25.00 & $7.1$ & \citet{Oudmaijer92} \\ 
60.00 & $2.5$ & \citet{Oudmaijer92} \\ 
65.00 & $1.7\pm0.1$ & \citet{Yamamura2010nu} \\ 
90.00 & $1.51\pm0.09$ & \citet{Yamamura2010nu} \\ 
100.00 & $0.9$ & \citet{Oudmaijer92} \\ 
450.00 & $<0.06$ & \citet{vdVeen94} \\ 
800.00 & $0.0016\pm0.0006$ & \citet{vdVeen94} \\ 
1100.00 & $0.0011\pm0.0004$ & \citet{vdVeen94} \\ 
1300.00 & $0.0011\pm0.0001$ & \citet{Garufi18} \\ 

  \hline
 \end{tabular}
\end{table}

\section{Stellar polarisation subtraction}\label{app:starpolsub}
Improper or incomplete removal of polarisation from the instrument, ISM or unresolved central star can result in $Q_\phi$ and $U_\phi$ artifacts which can be misinterpreted as disc structure.
To remove the stellar polarisation, we measure the fractional Q/I ($f_{\rm{Q}}$) and fractional U/I ($f_{\rm{U}}$). 
We then multiply this fractional polarisation with the intensity cube and subtract the new image from the Q and U frames, respectively (see e.g. \citealt{laws2020}). 
The instrumental polarisation will vary with the parallactic angle. Thus, stellar and instrumental polarisation are removed for each individual Stokes frame where the change in parallactic angle is minimal \citep{perrin2015}.
Since the stellar and instrumental polarisation are constant values of Q and U for a given observation cycle, the Stokes images are rotated into the $Q_\phi$ and $U_\phi$ frames. 
Any residual stellar or instrumental polarisation will form a quadrupole pattern and will be seen in the image (first column of Figure~\ref{fig:stellar_sub}).

Measuring $f_{\rm{Q}}$ and $f_{\rm{U}}$ can be complicated as we need to disentangle the stellar and instrumental polarisation signal from the unknown disc structure. 
We explored four different methods to measure the $f_{\rm{Q}}$ and $f_{\rm{U}}$ with the resulting mean-combined $Q_\phi$ and $U_\phi$ images shown in Figure~\ref{fig:stellar_sub}. 
We demonstrate these methods with the $J$-band data only. The $H$-band data showed similar results. 

\subsection*{Method 1}
This method is used in the GPI Data Reduction Pipeline (DRP) v1.5. All of the Q, U, and I counts within the coronagraphic spot ($<6\,$pix) are used and the calculated $f_{\rm{Q}}$ and $f_{\rm{U}}$ values are based on the summed Q, U, and I values. 
The resultant image (first column of Figure~\ref{fig:stellar_sub}) has a strong quadrupole signal, highlighting that this method is insufficient to remove the stellar and instrumental value. 
\citet{laws2020} drew similar conclusions: using the light within the coronagraphic spot does not result in the best subtraction of the stellar polarisation. 

\subsection*{Method 2}
We employed an algorithm that utilises the full field of view of the image while masking out individual pixels of the image where the disc is bright ($f_{\rm{Q}}$ or $f_{\rm{U}}$ $>0.05$) or where the signal-to-noise in the pixel is low. 
All the non-masked pixels are summed, giving summed values of Q, U, and I from which to calculate $f_{\rm{Q}}$ and $f_{\rm{U}}$.
This is the main method employed by the G-LIGHTS team (Rich et al. 2021b, \textit{in prep}). 
The resulting $Q_\phi$ and $U_\phi$ images are shown in the second column of Figure~\ref{fig:stellar_sub}. 
The quadrupole pattern is not as strong as for Method~1 but it is still present in the $U_\phi$ image. 
Additionally, the $Q_\phi$ image appears to indicate the presence of scattering along the apparent disc minor axis. 

\subsection*{Method 3}
\citet{laws2020} summed the Q, U, and I intensities in a $70<r<80\,$pixel radius around the target star. 
The resultant mean-averaged $Q_\phi$ and $U_\phi$ images are shown in the third column of Figure~\ref{fig:stellar_sub}. 
These show a further reduction in quadrupole structure, though some is still visible in the $U_\phi$ image.
However, the negative $Q_\phi$ flux values along the apparent disc minor axis produced using this method are not expected to be real. 

\subsection*{Method 4}
\begin{figure*}
    \centering
    \includegraphics[width=0.8\textwidth]{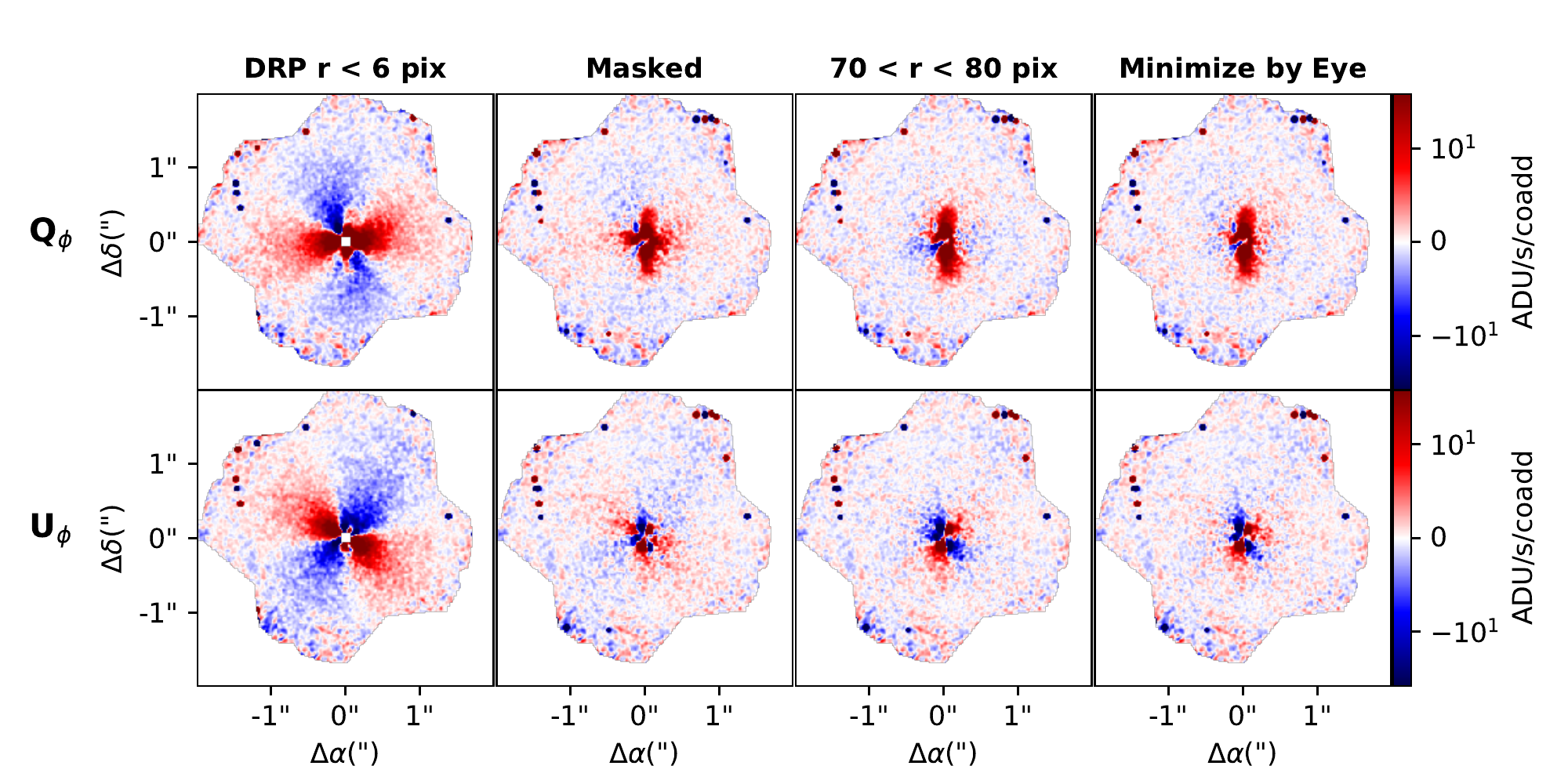}
    \caption{$J$-band mean-combined $Q_\phi$ (top row) and $U_\phi$ images (bottom row) for the four different stellar and instrumental polarisation removal techniques (see text for details). All images are plotted with the same symlog spread and smoothed with a Gaussian kernel to highlight the quadrupole pattern in the background seen in the first three columns.}
    \label{fig:stellar_sub}
\end{figure*}

This involves by-eye selection of the best values of $f_{\rm{Q}}$ and $f_{\rm{U}}$ to simultaneously minimise the quadrupole structure in the outer portions of the image and remove the positive/negative $Q_\phi$ flux along the apparent disc minor axis in the eight individual Stokes $Q_\phi$ and $U_\phi$ frames. 
The rationale here was to see whether the apparent minor axis structure found in the $Q_\phi$ image when using Method~2 (column 2 of Figure~\ref{fig:stellar_sub}) could solely be a result of residual stellar and instrumental polarisation. 
Ultimately, we can find a combination of $f_{\rm{Q}}$ and $f_{\rm{U}}$ for each of the eight Stokes frames which fully remove the residual quadrupole pattern and the minor axis feature with this method, as shown in the final column of Figure~\ref{fig:stellar_sub}. 
As this method results in no extraneous structures in the image, we use this by-eye minimisation of $Q_\phi$ and $U_\phi$ to remove stellar and instrumental polarisation herein.

\begin{figure}
    \centering
    \includegraphics[width=0.4\textwidth]{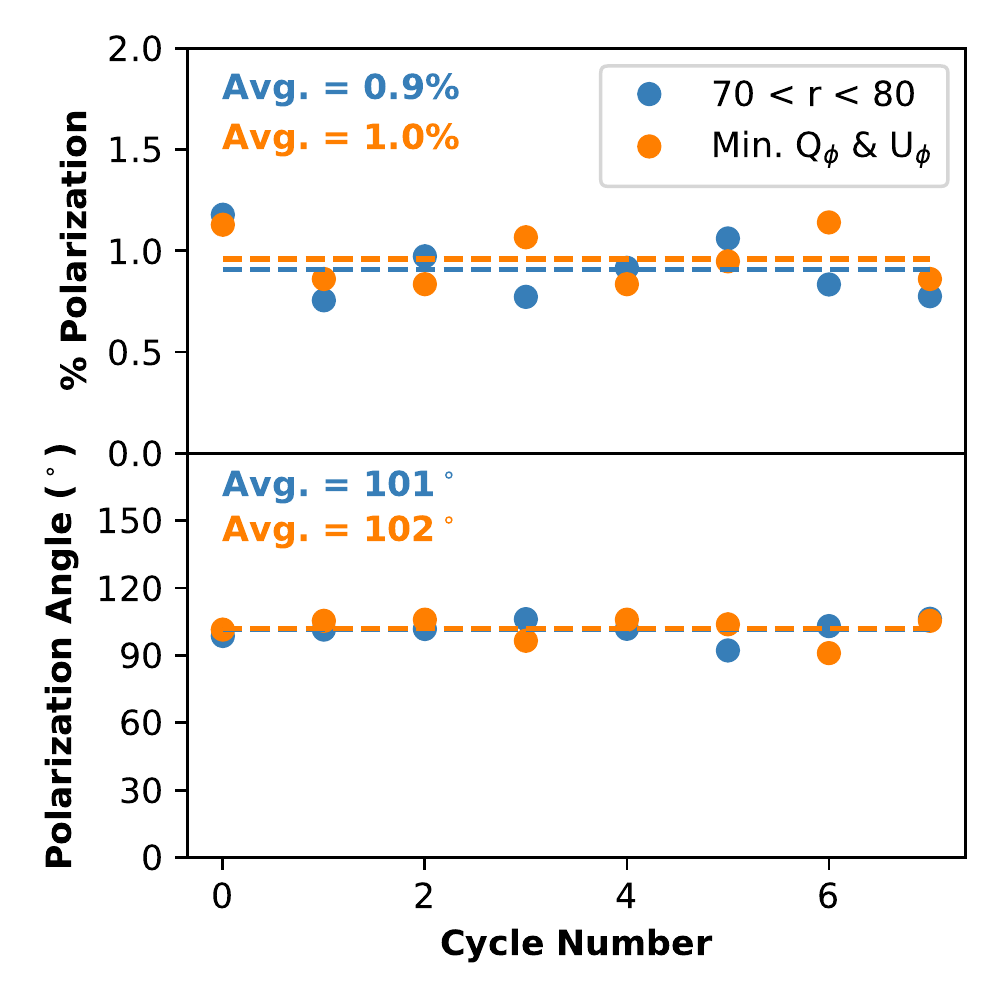}
    \caption{Percent polarisation (top panel) and polarisation angle (lower panel) of stellar and instrumental polarisation removed for each of the eight cycles of observations (method three is shown in blue; method four in orange). The average values for each of the methods are shown as dashed lines and the average values are written in the upper left of the panel.}
    \label{fig:stellar_removed}
\end{figure}

In Figure \ref{fig:stellar_removed}, we plot the stellar and instrumental polarisation angle and the percent polarisation for each of the eight Stokes cycles using Method~3 and Method~4. 
We measure an average percent polarisation of $1.0$ percent and an average polarisation angle of $102^{\circ}$. 
We note that both the polarisation angle and percent polarisation have strong deviations from the average with Method~4 resulting in standard deviations of $0.12\,$per cent and $5.1^{\circ}$, respectively. 
We also see that the average percent polarisation and polarisation angle for both methods do not differ wildly ($0.1\,$per cent and $1^{\circ}$) while some individual Stokes frames have very different polarisation values (e.g. Stokes cycle 3 and 6: Figure~\ref{fig:stellar_removed}).
Deviations from the average are expected as the magnitude and direction of the instrument polarisation changes as a function of parallactic angle. 
However, for continuous observations such as these, the stellar and instrumental polarisation should follow a sinusoidal function while the values shown in Figure~\ref{fig:stellar_removed} do not. 

One potential explanation for the change in variation is related to HD~145718 photometric variability. 
While this effect has not been studied in great detail to the best of our knowledge, dust from circumstellar occultation events should have a polarisation signal from surface scattering which would vary with orbital phase. 
Indeed, \citet{perrin2015} concluded from their instrumental polarisation measurements of $\beta$~Pic that GPI should have a polarisation accuracy of $0.1$ percent, slightly below the $0.12$ percent standard deviation we measure for HD~145718.
However, since there has been no systematic modeling of the instrument polarisation for GPI and our observations do not trace a sufficiently long timescale to allow us to undertake a similar analysis for HD~145718 as \citep{perrin2015} did for $\beta$~Pic, we cannot definitively conclude whether we are, in fact, observing polarisation variation from dust obscuration events. 

\section{Re-estimating stellar luminosity and visual extinction using bright epoch photometry}\label{app:starparam}
We followed \citet{Fairlamb15} and computed a grid of reddened \citet{Castelli03} model atmospheres with $T_{\rm{eff}}=7750$, $8000$, and $8250\,$K; $\log(g)=4.37$; and $A_{\rm{V}}$ ranging from $0.50$ to $2.50$ in steps of $0.01\,$mag. The value of $\log(g)$ makes little difference to the fit so its value is not changed. Each reddened model was fit to the \citet{Lazareff2017pv} $BVRI$ photometry and the best-fitting model was identified from the minimum of all the $\chi^{2}$ values. This procedure was run twice: once using the \citet{Cardelli89} reddening law with total-to-selective extinction, $R_{\rm{V}}=3.1$ and once using $R_{\rm{V}}=5.0$. We also re-ran the fitting to the fainter epoch photometry from \citet{Vieira03} to highlight the differences we observe. In each case, we follow \citet{Wichittanakom20} and use the models providing $\chi^{2}$ values twice the minimum-$\chi^{2}$ to estimate the uncertainty in $A_{\rm{V}}$.

\begin{figure}
    \centering
    \includegraphics[trim=0.1cm 1.20cm 0.cm 0.0cm, clip=true,width=0.44\textwidth]{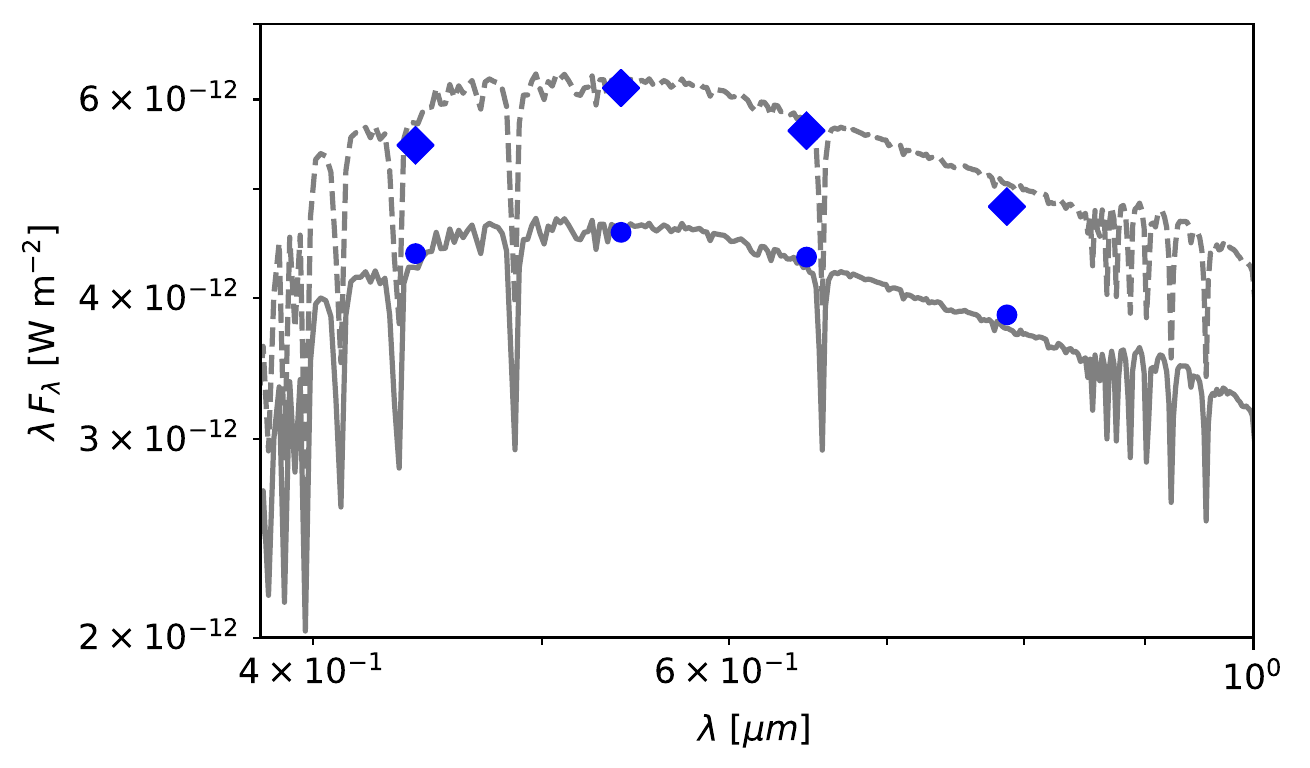}\\
    \includegraphics[trim=0.cm 0.cm 0.1cm 0.0cm, clip=true,width=0.44\textwidth]{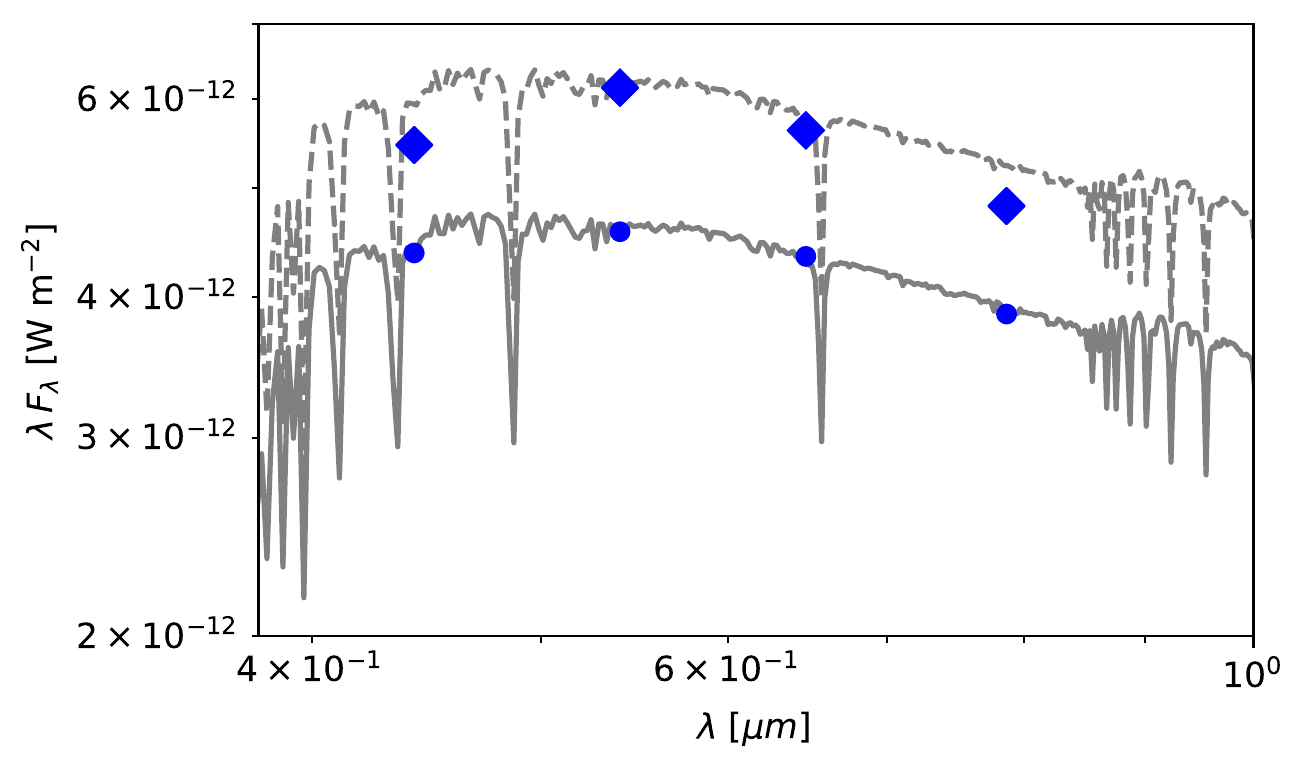}
    \caption{Comparison of \citet{Castelli03} model atmospheres with two epochs of previously published $BVRI$ photometry for HD~145718. The solid lines show our best fits to the \citet{Vieira03} photometry (filled circles) while dashed lines show our best fit to the \citet{Lazareff2017pv} photometry (filled diamonds). In the top and bottom plots, we show the fits adopting $R_{\rm{V}}=3.1$ and $R_{\rm{V}}=5.0$, respectively (see Table~\ref{tab:atmosphere_fits}).}
    \label{fig:atmosphere_fits}
\end{figure}

The reddened model atmosphere providing the best fit to the $BVRI$ photometry was then scaled to the $V$-band photometry. This scaling factor - which accounts for fitting models of surface flux to photometry - corresponds to $(d/R_{\star})^{2}$. We use this and the Gaia DR2 stellar distance to estimate the stellar radius, $R_{\star}$. The quoted uncertainties on $R_{\star}$ take into account the uncertainties on $A_{\rm{V}}$ and on $d$. Finally, we use the Stefan-Boltzmann law to calculate $L_{\star}$ from $T_{\rm{eff}}$ and $R_{\star}$, assuming an effective temperature for the Sun of $5771.8\,$K \citep{Mamajek12}. Our results are presented in Table~\ref{tab:atmosphere_fits} and displayed in Figure~\ref{fig:atmosphere_fits}.

The models reddened using $R_{\rm{V}}=3.1$ provide an improved fit to the \citet{Lazareff2017pv} photometry over those reddened using $R_{\rm{V}}=5.0$. Conversely, a better fit to the \citet{Vieira03} photometry is provided by the models reddened using $R_{\rm{V}}=5.0$. 

\begin{table}
    \centering
    \caption{Results from fitting \citet{Castelli03} stellar atmospheres with to $BVRI$ photometry. Column 1: the source of the photometry; column 2: the adopted total-to-selective extinction; columns 3 and 4: the best-fit $A_{\rm{V}}$ and $R_{\star}$, assuming $d=152.5\,$pc; column 5: inferred $L_{\star}$.} \label{tab:atmosphere_fits}
    \begin{tabular}{lcccc}
    \hline
    Photometry source & $R_{\rm{V}}$ & $A_{\rm{V}}$ & $R_{\star}$   & $L_{\star}$ \\
               & (mag)        & (mag)        & (R$_{\odot}$) & ($L_{\odot}$) \\ 
    (1)        & (2)          & (3)          & (4)           & (5) \\
    \hline
    \citet{Vieira03}       & $3.1$ & $0.89^{+0.31}_{-0.05}$ & $1.70^{+0.08}_{-0.07}$ & $10.7^{+2.6}_{-2.0}$ \\
    \citet{Vieira03}       & $5.0$ & $1.22^{+0.04}_{-0.03}$ & $1.98^{+0.07}_{-0.08}$ & $14.5^{+3.1}_{-2.7}$ \\
    \citet{Lazareff2017pv} & $3.1$ & $0.89^{+0.34}_{-0.08}$ & $1.97^{+0.12}_{-0.11}$ & $14.3^{+3.9}_{-3.1}$ \\
    \citet{Lazareff2017pv} & $5.0$ & $1.21^{+0.52}_{-0.21}$ & $2.28^{+0.35}_{-0.25}$ & $19.2^{+9.7}_{-5.8}$ \\
    \hline
    \end{tabular}
\end{table}

\bsp	
\label{lastpage}
\end{document}